\documentclass[11pt,a4paper]{article}
\usepackage{jheppub}
\usepackage{mathtools,bm}
\usepackage{booktabs,tabularx,array}
\usepackage[caption=false]{subfig}
\usepackage{microtype}
\hypersetup{
 pdftitle={Density-driven reversal of flavour-entanglement anisotropy in the D3-D7 holographic conductor},
 pdfauthor={Fuminori Okabayashi}
}
\newcommand{\Ncal}{\mathcal{N}_7}
\newcommand{\cT}{\mathcal T}
\newcommand{\cD}{\mathfrak D}
\newcommand{\cM}{\mathfrak M}
\newcommand{\cO}{\mathcal O}
\newcommand{\Tstar}{T_*}
\newcommand{\Vol}{\operatorname{Vol}}

\newcommand{\ii}{\mathrm{i}}

\title{Density-driven reversal of flavour-entanglement\\
anisotropy in the D3--D7 holographic conductor}
\author{Fuminori Okabayashi}
\affiliation{Department of Physics, Chuo University,\\
1-13-27 Kasuga, Bunkyo-ku, Tokyo 112-8551, Japan}
\emailAdd{fumi.okabayashi@gmail.com}

\abstract{
We study the leading probe-flavour contribution to equal-time spatial
entanglement entropy in a homogeneous $3+1$-dimensional strongly coupled
conductor at finite temperature and charge density, realised holographically
by the D3--D7 system and driven by a constant DC electric field.  We compare
slab-shaped subregions that are extended in two spatial directions and have
finite width parallel or transverse to the field.  The difference between the
corresponding entropy corrections isolates a homogeneous, spatially traceless
response and exhibits a density-driven sign reversal.
Over the investigated parameter range, the crossover density increases with the electric-field magnitude and
decreases with the strip width.  Mutual information shows the same qualitative
reversal and provides an ultraviolet-finite check.  We trace the effect to a
radial sign competition in the full Dirac--Born--Infeld stress anisotropy.
The reversal is a smooth response crossover rather than a thermodynamic phase
transition.
}


\begin{document}
\maketitle
\flushbottom

\section{Introduction}

Entanglement in non-equilibrium steady states (NESSs) is difficult to
characterise even when the steady state itself is under analytic control.
We ask how a conserved charge density modifies the orientation dependence of
equal-time spatial entanglement in a current-carrying state.  To address this
question, we study a homogeneous $3+1$-dimensional conductor at bath
temperature $T$, with a spatially uniform flavour charge density $d$, driven
by a DC electric field $E\hat x$ and carrying a steady current $J\hat x$.
The holographic description allows us to study a strongly coupled conductor
in three spatial dimensions, where the orientation of an extended entangling
region provides an independent probe of the driven state.

This boundary system is realised holographically by the D3--D7 conductor,
with $N_c$ colours and $N_f$ massless fundamental flavours.
For $N_f\ll N_c$, the flavour fields form an $\cO(N_fN_c)$ sector coupled
to the $\cO(N_c^2)$ adjoint plasma and are treated in the probe limit.
The electric field induces a non-linear current and continuous energy transfer
from the flavour sector to the adjoint bath
\cite{KarchOBannon2007,KarchOBannonThompson2009}.  The
Dirac--Born--Infeld (DBI) reality condition selects a radial shell
\cite{KarchOBannon2007}, commonly called the singular shell.  In the
open-string metric (OSM), this surface is a worldvolume horizon, also called
the effective horizon for probe fluctuations, governing their causal
propagation and effective temperature
\cite{KimShockTarrio2011,NakamuraOoguri2013,HoshinoNakamura2015}.

Holographic entanglement entropy is computed from extremal surfaces in the
closed-string geometry through the Ryu--Takayanagi (RT) and covariant
Hubeny--Rangamani--Takayanagi (HRT) prescriptions
\cite{RyuTakayanagi2006,HubenyRangamaniTakayanagi2007}.  At leading order in
the probe expansion, Chang and Karch expressed the flavour correction as the
first variation of the RT area sourced by the linearised closed-string
response to the probe stress tensor \cite{ChangKarch2014}.  In suitable
static Euclidean settings, the Karch--Uhlemann probe-action construction gives
the same correction without explicitly solving for the metric response
\cite{KarchUhlemann2014}.

A closely related probe-brane calculation analysed entanglement-entropy
production in the electrically driven D3--D7 system at zero charge density,
using linearised closed-string backreaction and the non-linear DBI current
\cite{OBannonProbstRodgersUhlemann2017}.
The observable studied here is complementary: the stationary orientation
difference at finite density, rather than the common time-dependent heating
response.

Ref.~\cite{BanerjeeEtAl2021} proposed to compute flavour-sector entanglement
entropy by applying the RT area prescription to the open-string metric.
This appealing conjecture is motivated by the emergent horizon and effective
thermality of probe fluctuations.  The original study found suggestive
thermodynamic relations while recognising that the identification required
further justification.  However, the static comparison in
ref.~\cite{OkabayashiPaperI} with the controlled finite-density result
\cite{ChangKarchUhlemann2014}, together with the driven comparisons in
section~\ref{sec:osm-comparison}, shows that the bare OSM areas tested do
not reproduce the controlled leading flavour contribution to boundary
entanglement entropy.  The proposed identification is therefore not generally
valid as a bare-area prescription.  This conclusion does not exclude a
different open-string functional derived from a boundary replica construction.

Orientation-dependent holographic entanglement has been studied both in
boosted black branes \cite{MishraSingh2016} and in intrinsically anisotropic
backgrounds.  Q-lattice models exhibit subsystem-size-dependent changes in
the angular ordering of mutual information \cite{LiuNiuWu2019}.
In equilibrium p-wave phases, the difference between parallel and transverse
strip entropies has been used as an anisotropy order parameter
\cite{ParkEtAl2022}.  An emergent isotropic point of the near-horizon
geometry has also been studied \cite{ChenEtAl2024}; this need not imply
isotropy of the full finite-size entanglement entropy.
Multipartite entanglement measures have also been used to probe topological
transitions and anisotropic infrared behaviour in holographic Weyl semimetals
\cite{ChenEtAl2026}.
In those settings the anisotropy is a property of the leading closed-string geometry and may be tied
to equilibrium phase structure.  Here the adjoint bath remains isotropic, while the anisotropy appears
only in the order-$N_fN_c$ flavour correction sourced by the driven D7 brane.  The density-driven sign
reversal studied below is therefore a probe-sector response crossover rather than an
anisotropic-background RT effect or a symmetry-restoring transition.

A complication arises when the backreaction of the driven sector is retained.
At leading probe order, the adjoint plasma acts as an effective heat sink:
energy and, at nonzero charge density, momentum are continuously transferred
to it \cite{KarchOBannonThompson2009}.  Once its response is retained,
their accumulation produces time-dependent closed-string contributions,
including the heating response isolated in
ref.~\cite{OBannonProbstRodgersUhlemann2017}.  Maintaining a fully
stationary backreacted state would require an additional prescription for
removing these inputs; such a prescription is not part of the standard probe
solution.  Absolute scalar-sector entropy coefficients can consequently
depend on how the state is completed.

We avoid this ambiguity by comparing
two strip orientations.\footnote{We use `strip' following common terminology in holographic
entanglement studies. Here it denotes a three-dimensional spatial region
of finite width in one direction and extended in the other two. We also
use `slab' for the same region to make this geometry explicit for
readers from non-equilibrium statistical physics.}  For strips whose finite direction is parallel or transverse to the field, we
denote the corresponding leading D7-induced corrections by $\Delta S_{\parallel}$ and
$\Delta S_{\perp}$.  Their difference $\Delta S_{\parallel}-\Delta S_{\perp}$
projects onto the homogeneous, spatially traceless response sourced by
$T^x{}_x-T^y{}_y$, the longitudinal-minus-transverse pressure contrast of
the effective bulk D7 source.  This radial source is not directly the boundary
stress-tensor expectation value.  On the isotropic unbackreacted AdS$_5$--Schwarzschild background this external
tensor channel closes at linear order, while the common scalar/trace response cancels from the
orientation difference.  We compute only the D7-sourced leading flavour correction.  An additional
reservoir carrying its own order-$N_fN_c$ anisotropic stress would constitute a separate tensor source
and is not included here.

The resulting orientation difference exhibits a density-driven sign reversal.
We are not aware of a previous calculation of such a reversal of the leading probe-flavour
orientation response in a current-carrying holographic conductor.  At fixed strip width $\ell$ and
dimensionless field $e=Ez_h^2$, with $E$ the applied electric field and $z_h$ the background-horizon
radius, the orientation difference changes sign at a dimensionless density $\delta_c(e,\ell)$.  The
crossover moves to larger density as the electric-field magnitude increases and to lower density as the region probes further
into the infrared.  The sign reversal is also present in mutual information, where ultraviolet (UV) divergences
cancel.  We trace these trends to the radial sign of the full DBI stress anisotropy, whose local
crossover density is a monotonic function of the radial position and the electric-field magnitude.
The reversal line is reproduced by three independent numerical implementations of the same
controlled backreaction.

Non-equilibrium entanglement depends on the occupation pattern, the dynamics
of correlations and the choice of subsystem.  Certain partitioning states in
conformal field theory (CFT) admit a boosted-thermal description
\cite{HoogeveenDoyon2015}, while temperature-biased free-fermion chains
exhibit logarithmic violations of the mutual-information area law
\cite{EislerZimboras2014}.  At zero reservoir temperature, coherent
conductors with partially reflecting contacts or a local scatterer can
have extensive subsystem entropy with logarithmic corrections
\cite{Ribeiro2017,FraenkelGoldstein2021}.  Coherent transmission and
reflection also sustain extensive long-range entanglement between regions
with mirror overlap across a scatterer \cite{FraenkelGoldstein2023}.
The coherent-conductor examples identify the partitioning of unequally
occupied incoming modes, rather than current flow alone, as a recurring
mechanism.

In a disordered wire, Hakoshima and Shimizu found a quasi-volume enhancement
for a region inside the wire, requiring both far-from-equilibrium bias and
coherent multiple scattering \cite{HakoshimaShimizu2019}.  The survival
and spatial extent of coherence also matter in diffusive systems:
noninteracting boundary-driven circuits exhibit volume-law entanglement,
whereas chaotic interactions can promote local equilibration and short-range
entanglement; a three-dimensional Anderson conductor exhibits extensive
mutual information \cite{GullansHuse2019}.  A subsequent study of the
Anderson transition found extensive scaling of mutual coherence, a fermionic
entanglement witness, up to the localisation critical point and area-law
scaling in the localised phase \cite{GullansHuseLocalization2019}.
The common role of coherent mode partitioning is therefore informative,
although neither scattering strength nor the presence of a current alone
classifies the resulting correlations.

HRT calculations have also tracked entanglement during the
formation of partitioning NESSs.  In AdS$_3$/CFT$_2$, joining heat baths at
different temperatures produces an expanding steady-state region for which
the time dependence of entanglement entropy and mutual information was
computed \cite{ErdmengerEtAl2017NESS}.  In $3+1$-dimensional strongly coupled
$\mathcal N=4$ super-Yang--Mills theory, partitioning protocols with temperature
and chemical-potential imbalance produce a steady region with non-trivial
charge-density structure, while HRT entropies track the passage of shock and
rarefaction waves \cite{EckerErdmengerVanDerSchee2021}.  In that calculation
the gauge-field backreaction was neglected, so the HRT entropy does not
directly measure the charge-sector correction.  A more closely related
electric-field study considered holographic entanglement in a homogeneous
momentum-relaxing DC setup: the electric-field correction to strip entanglement
entropy vanishes at linear order, whereas a rotated sharp-wedge response is proportional to the thermoelectric
conductivity \cite{KimParkLeeAhn2019}.  These partitioning and
momentum-relaxing studies complement the D3--D7 entanglement-rate calculation
discussed above.

Against this broader landscape, the orientation difference isolates a
density-tunable tensor response and relates its reversal to the radial
competition in the DBI source, without requiring the absolute large-region
scaling of the flavour entropy.

The remainder of the paper is organised as follows.  Section~\ref{sec:setup}
introduces the D3--D7 steady state and conventions.  Section~\ref{sec:probeEE}
derives the tensor-channel Green-function representation,
sections~\ref{sec:strip} and \ref{sec:mi} present the strip and
mutual-information results, section~\ref{sec:numerics} summarises the
numerical validation, and sections~\ref{sec:discussion} and
\ref{sec:conclusion} discuss the interpretation and conclusions.
Appendix~\ref{app:source} derives the DBI tensor source,
appendix~\ref{app:norm} fixes the strip normalisation,
appendices~\ref{app:MIshift} and \ref{app:numerics} give the mutual-information transition shift and
numerical details, and appendix~\ref{app:scalar-volume} records the
auxiliary scalar response and its temperature matching.

\section{D3--D7 conductor and conventions}\label{sec:setup}

The bulk setup is dual to large-$N_c$, strongly coupled
$\mathcal N=4$ $SU(N_c)$ super-Yang--Mills theory coupled to
$N_f\ll N_c$ massless $\mathcal N=2$ fundamental hypermultiplets
\cite{Maldacena1998,KarchKatz2002}.  The
external electric field sources the spatial component of the flavour
$U(1)_B$ current, and $d$ denotes the corresponding charge density.
We set the AdS radius $L=1$ in the working metric and use the rescaled
worldvolume gauge field $A=(2\pi\alpha'/L^2)A_{\mathrm{phys}}$ throughout,
where $\alpha'$ is the Regge slope.  In particular,
$E=(2\pi\alpha'/L^2)E_{\mathrm{phys}}$; the same constant factor is
absorbed into the field strength.

The AdS$_5$--Schwarzschild metric is
\begin{equation}
 ds^2=\frac{1}{z^2}\left[
 -f(z)\,dt^2+dx^2+dy^2+dw^2+\frac{dz^2}{f(z)}
 \right],
 \qquad
 f(z)=1-\left(\frac{z}{z_h}\right)^4 .
\end{equation}
The black-brane Hawking temperature
$T=(\pi z_h)^{-1}$ is identified with the temperature of the adjoint plasma
that acts as the heat bath.  The massless D7 wraps the AdS$_5$ directions and
an equatorial $S^3\subset S^5$.  After integrating over the internal sphere,
the DBI action is
\begin{equation}
 S_{\mathrm{D7}}
 =-\Ncal\int d^4x\,dz\,
 \sqrt{-\det(g_{ab}+F_{ab})},
 \qquad F=dA .
\end{equation}
Here $g_{ab}$ is the external five-dimensional block of the induced D7
worldvolume metric; the determinant and the indices $a,b$ refer to the
$(t,x,y,w,z)$ directions after integration over the internal $S^3$.
For reference, the normalisation and the conversion to the dimensionless
electric field are
\begin{equation}
 \Ncal=N_fT_7L^8\Vol(S^3)
 =\frac{\lambda N_fN_c}{(2\pi)^4},
 \qquad
 e=\frac{2\pi\alpha'}{L^2}E_{\mathrm{phys}}z_h^2 .
\end{equation}
Here $T_7$ is the D7-brane tension, $g_{\mathrm{YM}}$ is the Yang--Mills
coupling, $\lambda=g_{\mathrm{YM}}^2N_c$ is the 't~Hooft coupling and
$E_{\mathrm{phys}}$ is the physical electric field.
In these conventions, $e=Ez_h^2$.

The stationary gauge field is
\begin{equation}
 A_x=-Et+a_x(z),\qquad A_t=A_t(z).
\end{equation}
The signed density and current are defined by the UV boundary variation:
\begin{equation*}
 \left.\delta S_{\mathrm{D7}}^{\mathrm{on\mbox{-}shell}}\right|_{\mathrm{UV}}
 =\int d^4x\,[d\,\delta A_t^{(0)}+J\,\delta A_x^{(0)}].
\end{equation*}
Here $A_\mu^{(0)}$ denotes the boundary value of the gauge field.
Because $z$ increases into the bulk, this fixes
$d=-\partial\mathcal L_{\mathrm{D7}}/\partial A_t'$ and
$J=-\partial\mathcal L_{\mathrm{D7}}/\partial a_x'$, with primes here
denoting $z$ derivatives.  In the rescaled gauge-field convention these are
the conserved charge density and electric current.  We define
\begin{equation}
 u=\frac{z}{z_h},\qquad
 \delta=\frac{dz_h^3}{\Ncal},\qquad
 j=\frac{Jz_h^3}{\Ncal},\qquad
 s=\sqrt{1+e^2}.
\end{equation}
We also use dimensionless boundary coordinates
$\bar x^\mu=x^\mu/z_h$, in particular $\bar t=t/z_h$ and $\bar x=x/z_h$.
Because the tensor observables depend only on $e^2$ and $\delta^2$, we
take $e,\delta\geq0$ without loss of generality.  The DBI reality
condition fixes the current at the worldvolume horizon.\footnote{The conventional real-action prescription for the DBI
conductor was introduced in ref.~\cite{KarchOBannon2007} and uses the
square-root structure of the action.  Ref.~\cite{IshigakiNakamuraTakasan2024}
formulates algebraic patchwork conditions for more general non-linear models
and conjectures their equivalence to regularity, with the appropriate
boundary conditions and checks in the examples studied there.}
The horizon position is
\begin{equation}
 u_*=(1+e^2)^{-1/4}=s^{-1/2},
\end{equation}
and the current obeys
\begin{equation}
 j^2=e^2\left(s+\frac{\delta^2}{s^2}\right).
\end{equation}

The two positive contributions to $j^2/e^2$ have distinct physical origins
\cite{KarchOBannon2007,HoshinoNakamura2015}.  The term $s$ is the
charge-conjugation-symmetric sector conventionally associated with pair
creation, including field-induced pairs, whereas $\delta^2/s^2$ is the
pre-existing net-density (or doped) sector.  This decomposition is a useful
physical interpretation of the squared current, not a quasiparticle-level sum
of independently additive currents.  Density changes the relative weight of
these sectors but does not itself select a spatial direction; the anisotropy
direction is fixed by the electric field.

For the radial first integrals, define the dimensionless profiles
$\widetilde A_t(u)=z_h A_t(z_hu)$ and
$\widetilde a_x(u)=z_h a_x(z_hu)$, and suppress the tildes below.  Primes in
the next two equations denote derivatives with respect to $u$, and
$f(u)=1-u^4$.  Eliminating the radial gauge-field derivatives in favour of
$d$ and $J$ then gives
\begin{equation}
 A_t'=-\delta u\sqrt{\mathcal X},
 \qquad
 a_x'=\frac{ju}{f}\sqrt{\mathcal X},
\end{equation}
where
\begin{equation}
 \mathcal X(u;e,\delta)=
 \frac{s^2(1+s u^2)}
 {(1+s u^2)(s^2+\delta^2u^6)+s^2(s^2-1)u^4}.
 \label{eq:Xregular}
\end{equation}
For nonzero field the physical current branch has $ej>0$.  In dimensionless
coordinates its radial energy flux is $-T^u{}_{\bar t}=\Ncal eju^5>0$, directed towards the
interior.  For ingoing Eddington--Finkelstein (EF) time
$v=\bar t-\int^u d\hat u/f(\hat u)$, the metric has
cross term $-2\,dv\,du/u^2$ and
$F_{u\bar x}=(ju\sqrt{\mathcal X}-e)/f$.
Thus $T^u{}_v=T^u{}_{\bar t}$ has the same ingoing sign.

\begin{figure}[t]
 \centering
 \includegraphics[width=0.94\textwidth]{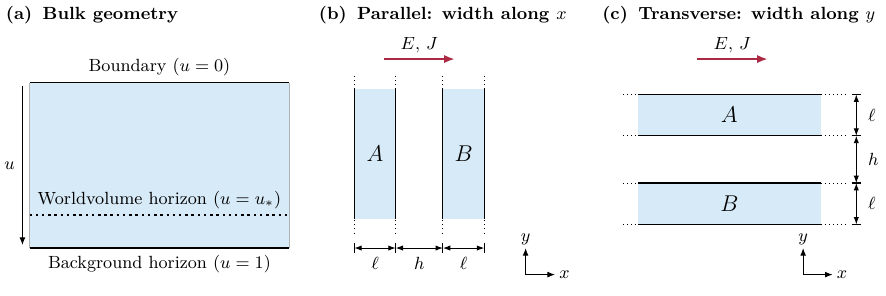}
 \caption{Geometry and subsystem orientations (schematic).
 (a) Radial view of the bulk, showing the boundary at $u=0$, the D7
 worldvolume horizon at $u=u_*$ and the background horizon at $u=1$.
 The massless D7 fills the displayed external directions; its internal
 $S^3$ is not shown.
 (b,c) Boundary $x$--$y$ sections of two regions $A$ and $B$, each of
 finite width $\ell$ and separated by a gap $h$. The width and separation
 are along $x$ for the parallel orientation and along $y$ for the
 transverse orientation. The electric field and current point along $+x$
 in both cases. Dotted continuations indicate the extended direction in
 the displayed plane; both regions also extend along the suppressed
 $w$ direction. The single-strip calculation uses $S(A)$ or $S(B)$;
 the separation $h$ enters only the two-region mutual information
 $I(A:B)=S(A)+S(B)-S(A\cup B)$.}
 \label{fig:setup-schematic}
\end{figure}

The DBI dependence on $e$ and $\delta$ is retained non-perturbatively, while
the closed-string response is computed only at leading order in $N_f/N_c$.
The large-$N_c$ counting is at fixed large $\lambda$.  With
the five-dimensional Newton constant $G_5=\pi/(2N_c^2)$ in units $L=1$, the metric expansion parameter is
$\kappa=8\pi G_5\Ncal=\lambda N_f/(4\pi^2N_c)\ll1$; the response must
also remain small throughout the radial region probed by the surface.
From the boundary transport viewpoint, $T$, $E$ and $d$ are control
parameters, while $J$ is the steady response fixed by worldvolume regularity.
Here the `flavour contribution' denotes the $\cO(N_fN_c)$ correction induced
by the flavour sector to the equal-time spatial entanglement entropy of the
full boundary theory; it is not a species-resolved entropy of the D7 degrees
of freedom alone.
The geometry and orientation conventions are summarised in figure~\ref{fig:setup-schematic}; in
particular, the D7 worldvolume horizon $u_*$ and background horizon $u=1$ are distinct.

\section{Leading probe entanglement without constructing the full backreaction}\label{sec:probeEE}

\subsection{Probe-order Green-function representation}

For a boundary region $A$, the physical metric perturbation is counted as
\begin{equation}
 g_{MN}=g^{(0)}_{MN}+h^{(1)}_{MN}
 +\cO\!\left((N_f/N_c)^2\right),
 \qquad h^{(1)}_{MN}=\cO(N_f/N_c),
\end{equation}
and the entropy expansion is
\begin{equation}
 S_A=S_A^{(0)}+S_{A,\mathrm{fl}}^{(1)}+\cO(N_f^2),
 \qquad S_A^{(0)}=\cO(N_c^2),
 \qquad S_{A,\mathrm{fl}}^{(1)}=\cO(N_fN_c).
\end{equation}
At this order, the physical D7 source already carries its $N_f$ normalisation, and
\begin{equation}
 h^{(1)}_{MN}(X)=8\pi G_5\int dX'\,G^{(0)}_{R,MN|PQ}(X,X')T_{\mathrm{D7}}^{PQ}(X'),
 \label{eq:hgreen}
\end{equation}
where $G_5$ is the five-dimensional Newton constant, $X$ is a five-dimensional bulk point, and
$T_{\mathrm{D7}}^{PQ}$ is the effective five-dimensional D7 stress tensor after internal-space
integration.  We use $dX'\equiv d^5X'$ and absorb the invariant measure factor
$\sqrt{-g^{(0)}(X')}$ into the definition of
$G^{(0)}_{R,MN|PQ}(X,X')$.  The retarded propagator is evaluated in the unbackreacted geometry.
Replacing it by the propagator of
$g^{(0)}+h^{(1)}$ would affect the result only at $\cO((N_f/N_c)^2)$.

According to the covariant HRT prescription
\cite{HubenyRangamaniTakayanagi2007}, the relevant extremal surface for the
present equal-time region lies on the time-reflection-symmetric slice of the
static adjoint background and reduces to the RT surface
\cite{RyuTakayanagi2006}.  In the five-dimensional effective description,
let $\Sigma_A^{(0)}$ denote the external three-dimensional part of this
unperturbed surface; the internal zero-mode reduction is justified in the
next subsection.  Its induced metric is $\gamma_{ab}^{(0)}$, its intrinsic
coordinates are $\sigma^a$, and its embedding is $X^M(\sigma)$.  Extremality removes the
first-order embedding variation, yielding
\begin{equation}
 S_{A,\mathrm{fl}}^{(1)}=
 \frac{1}{8G_5}\int_{\Sigma_A^{(0)}}d^3\sigma\,
 \sqrt{\gamma^{(0)}}\,\gamma_{(0)}^{ab}h^{(1)}_{MN}
 \partial_aX^M\partial_bX^N.
 \label{eq:areaVariation}
\end{equation}
Equations~\eqref{eq:hgreen} and \eqref{eq:areaVariation} give the Chang--Karch representation of the
leading probe contribution to the entropy \cite{ChangKarch2014}.  This Green-function representation is convenient but not
mandatory: direct solution of the tensor boundary-value problem provides an independent check.

\subsection{External tensor projection}

Here scalar, vector and tensor refer to representations of external spatial
rotations of the unperturbed isotropic geometry at zero boundary momentum.
The spatial trace and components with both indices in $\{t,u\}$ are scalars;
$h_{ti}$ and $h_{ui}$, with $i\in\{x,y,w\}$, are vectors; and the symmetric
traceless spatial perturbation is a tensor.  These sectors decouple at linear
order.  This classification is distinct from the internal $S^5$ harmonic
decomposition.

The residual symmetry in the $y$--$w$ plane gives the effective bulk D7
stress the spatial form $\mathrm{diag}(p_\parallel,p_\perp,p_\perp)$, where
$p_\parallel=T^x{}_x$ and $p_\perp=T^y{}_y=T^w{}_w$ depend on the radial
coordinate.  Its decomposition is
\[
 \mathrm{diag}(p_\parallel,p_\perp,p_\perp)
 =\frac{p_\parallel+2p_\perp}{3}\,\mathbf 1
 +(p_\parallel-p_\perp)
 \mathrm{diag}\left(\frac23,-\frac13,-\frac13\right).
\]
Thus $T^x{}_x-T^y{}_y$ is precisely the longitudinal-minus-transverse
pressure contrast sourcing the external traceless response.

The ten-dimensional RT surface wraps the full internal $S^5$.  Expand an external metric response in
internal scalar harmonics, $h_{\mu\nu}(x,\Omega)=\sum_I h_{\mu\nu}^{(I)}(x)Y_I(\Omega)$, where
$\Omega$ denotes the internal coordinates.  On the
unperturbed surface, the internal part of the first-order area measure is the round-sphere measure, and
\begin{equation}
 \int_{S^5}d\Omega_5\,\sqrt{g_{S^5}}\,Y_{I>0}(\Omega)=0.
\end{equation}
The $S^5$ integral in the area variation therefore projects the external metric
perturbation onto its scalar zero harmonic.  The linearised equations and the
boundary and horizon conditions preserve the internal isometries.  Within the
external spatially traceless tensor sector, the linearised operator is
therefore diagonal in the $S^5$ scalar harmonics; the zero mode is sourced by
the internal average of $T^x{}_x-T^y{}_y$.
Ref.~\cite{ChangKarch2014} established the corresponding reduction for the
Euclidean/static Green function.  That work also noted that infalling horizon
conditions spoil Hermiticity, so its mode argument need not apply to a
finite-frequency retarded propagator.  We do not assume such an extension
here: section~\ref{sec:retarded-static} shows that,
in the present homogeneous tensor channel, $G_R(\omega=\ii0^+)$ reduces to
the unique static kernel used below.

We work in radial gauge, $h_{uM}=0$.  The homogeneous external traceless
perturbation may be chosen as
\begin{equation}
 h^i{}_j=\Phi(u)\,\mathrm{diag}\left(1,-\frac12,-\frac12\right),
 \qquad i,j\in\{x,y,w\}.
 \label{eq:tensorAnsatz}
\end{equation}
Equivalently, $h^x{}_x=\Phi$ and $h^y{}_y=h^w{}_w=-\Phi/2$; all other components vanish within
this projected tensor channel, not in the full metric perturbation.
The same projection occurs in the area variation.  A common spatial-trace
perturbation contributes equally to the two rotated strips.  In radial gauge
the remaining vector components are $h_{ti}$, whose pullback vanishes on the
unperturbed equal-time surfaces; the first-order embedding variation vanishes
by extremality.  For the tensor ansatz, the difference of the two
half-surface area weights is
\[
 \left[\frac12\gamma_{(0)}^{ab}\delta\gamma_{ab}\right]_{
 \parallel-\perp}
 =-\frac34\left[1-\left(\frac{u}{u_t}\right)^6\right]\Phi(u),
\]
where $u_t$ is the unperturbed strip turning point.  The common scalar
response therefore cancels, leaving this tensor contribution; the complete
normalisation is given in appendix~\ref{app:norm}.

The D7 Wess--Zumino action
contains the pullback of the background Ramond--Ramond four-form potential
$C_4$.  For the worldvolume field strength
\begin{equation}
 F=-E\,dt\wedge dx+A_t'\,dz\wedge dt+a_x'\,dz\wedge dx,
\end{equation}
where primes denote derivatives with respect to $z$, one has $F\wedge F=0$.
The $C_4\wedge F\wedge F$ secondary source that can obstruct the simple
probe-stress-tensor formula is therefore absent for this ansatz
\cite{ChangKarch2014}.

\subsection{Full DBI tensor source and normalisation}

After internal-space integration and zero-mode projection, define the effective five-di\-men\-sion\-al
mixed-index D7 source entering the tensor Einstein equation,
\begin{equation}
 \cT_T=T^x{}_x-T^y{}_y,
 \qquad
 \widehat{\cT}_T=\frac{\cT_T}{\Ncal e^2}.
\end{equation}
The full DBI source, derived in appendix~\ref{app:source}, is
\begin{equation}
 \widehat{\cT}_T(u;e,\delta)=
 \frac{u^4\sqrt{\mathcal X(u)}}{s^2(1+s u^2)}
 \left[\delta^2u^2(1+s u^2)-s^2\right].
 \label{eq:tensorSource}
\end{equation}
Its UV expansion is $\widehat{\cT}_T=-u^4+\cO(u^6)$.  The D7 stress anisotropy is finite both at
$u=u_*$ and at the background horizon $u=1$; the worldvolume horizon is not an endpoint of the
closed-string tensor problem.

It is convenient to factor out the physical coefficient by writing
\begin{equation}
 \Phi(u)=\frac{32\pi G_5\Ncal e^2}{3}\,\varphi(u).
\end{equation}
The normalised tensor mode obeys
\begin{equation}
 \partial_u\left[\frac{1-u^4}{u^3}\varphi'(u)\right]
 =-\frac{\widehat{\cT}_T(u)}{u^5},
 \qquad \varphi(0)=0,
 \quad \varphi\ \hbox{regular at }u=1.
 \label{eq:tensorODE}
\end{equation}
The static kernel is the $\omega\to0$ limit of the retarded tensor response; equivalently, it is the
Green function for the unique boundary-Dirichlet, horizon-regular solution in this channel.  The
corresponding Green function is
\begin{equation}
 G_T(u,u')=\frac14\log\left[1-\min(u,u')^4\right],
\end{equation}
so that
\begin{equation}
 \varphi(u)=-\int_0^1du'\,G_T(u,u')\frac{\widehat{\cT}_T(u')}{u'^5}.
 \label{eq:phiGreen}
\end{equation}

\subsection{Retarded prescription and static limit}
\label{sec:retarded-static}

Although the unperturbed adjoint geometry is static at leading probe order, the D7 sector is in a
Lorentzian current-carrying steady state.  Denote the average metric perturbation on the two
Schwinger--Keldysh contour legs by $h_r=(h_1+h_2)/2$.  On the physical contour the D7 stress is the
same on both legs and therefore acts as an ordinary physical source.  At classical linear order the
mean metric perturbation consequently responds through the retarded graviton propagator,
\begin{equation}
 h_r=8\pi G_5\,G_R*T_{\mathrm{D7}}.
 \label{eq:retardedMeanField}
\end{equation}
Here $*$ denotes the spacetime convolution displayed explicitly in
eq.~\eqref{eq:hgreen}.  Only the retarded kernel contributes to this
classical mean response to equal physical sources on the two contour legs;
the Keldysh correlator describes fluctuations and is not assumed to vanish.
The infalling prescription for holographic retarded correlators
\cite{SonStarinets2002} and its thermal Schwinger--Keldysh formulation
\cite{HerzogSon2003} are consistent with this causal response.  The more
general contour prescription specifies real-time state preparation
\cite{SkenderisVanRees2009}.  For a stationary source prepared adiabatically
from the remote past, the convolution selects $G_R(\omega=\ii0^+)$.
A real-time replica derivation of HRT is itself formulated on a
Schwinger--Keldysh contour \cite{DongLewkowyczRangamani2016}; here we use
only the linearised HRT-area observable evaluated on the retarded mean metric.

In the homogeneous, spatially traceless channel considered here, the zero-frequency limit is
unobstructed.  A homogeneous solution of
\begin{equation}
 \mathcal L_T\phi
 \equiv
 \partial_u\left[\frac{1-u^4}{u^3}\partial_u\phi\right]=0
\end{equation}
obeys
\begin{equation}
 \frac{1-u^4}{u^3}\phi'=C.
\end{equation}
Here $C$ is an integration constant.
For $C\neq0$, the solution develops a logarithmic singularity at the background future horizon
$u=1$.  Horizon regularity therefore requires $C=0$, and the boundary Dirichlet condition
$\phi(0)=0$ then gives $\phi=0$.  Hence the static boundary-value problem has no zero mode and the
retarded limit is the unique real static kernel,
\begin{equation}
 G_R^T(\ii0^+;u,u')
 =G_T(u,u')
 =\frac14\log\left[1-\min(u,u')^4\right].
 \label{eq:retardedStaticCollapse}
\end{equation}
A stronger equivalence with an on-shell D7 replica-action formula would require additional control of replicated
probe saddles and is not needed for the results below.

\section{Strip anisotropy and the density-driven crossover}\label{sec:strip}

\begin{figure}[t]
 \centering
 \subfloat[Radial profiles of the normalised DBI tensor source at $e=0.5$.]{%
  \includegraphics[width=0.48\textwidth]{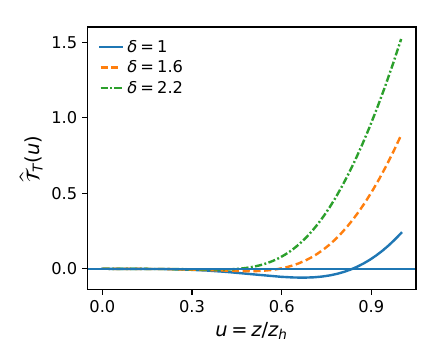}}\hfill
 \subfloat[The entropy difference between orientations at $e=0.5$ for several strip widths.]{%
  \includegraphics[width=0.48\textwidth]{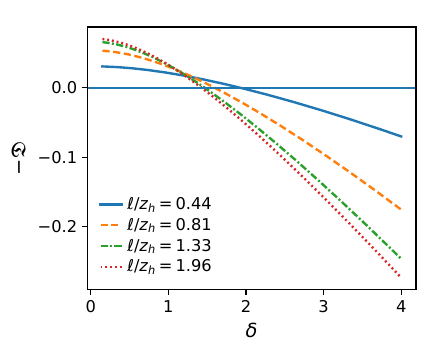}}
 \caption{Origin of the density-driven reversal.  The full DBI source changes its radial sign as the
 density is increased, and the RT/Green kernel converts this sign competition into a zero of the
 orientation-dependent entanglement correction.  Panel (b) plots $-\cD$, which is proportional to
 $\Delta S_{\parallel}-\Delta S_{\perp}$ according to the convention in
 eq.~\eqref{eq:entropyDifference}.  The entropy-sign curves use fixed-order quadrature of order
 $N=720$.  Colour and line style distinguish the curves.}
 \label{fig:source-and-sign}
\end{figure}

The boundary subregion is a slab of volume $V_2\ell$ in the three-dimensional
spatial theory: it is finite in one direction and extended in the other two.
The electric field points along $x$.  We call a strip \emph{parallel} when its finite direction is
$x$ and its two extended directions are $y,w$; a \emph{transverse} strip has finite direction $y$
and extends along $x,w$.  We regulate the two extended directions with the same area
$V_2=\int dy\,dw=\int dx\,dw$.
To determine the unperturbed strip, parametrise one radial half by the
finite-direction coordinate $\bar x^i(u)=x^i(u)/z_h$, with $i=x$ or $y$.
Its area density is proportional to
$u^{-3}\sqrt{f^{-1}+(\partial_u\bar x^i)^2}$.
Translation invariance in $\bar x^i$ gives a conserved conjugate momentum,
whose magnitude is fixed by the turning point to $u_t^{-3}$, and hence
\[
 \left|\partial_u\bar x^i(u)\right|
 =\frac{u^3}{\sqrt{f(u)}\sqrt{u_t^6-u^6}}.
\]
Integrating from the boundary to the turning point and adding the two halves
gives
\begin{equation}
 \frac{\ell}{2z_h}=\int_0^{u_t}du\,
 \frac{u^3}{\sqrt{1-u^4}\sqrt{u_t^6-u^6}}.
 \label{eq:stripWidth}
\end{equation}
For finite width, we evaluate the connected U-shaped background RT surface.
In the planar AdS$_5$--Schwarzschild geometry this smooth branch exists for
every finite width and approaches the horizon as $\ell\to\infty$.  A
homology-completed disconnected comparison configuration, consisting of two
vertical sheets and a horizon segment, can also be formed, but the connected
surface has lower area for all widths; see section~3.2 and figure~2 of
ref.~\cite{BahEtAl2009}.  Hence
there is no single-strip topology transition in the present background.
At fixed dimensionless parameters within the controlled probe regime, the
background entropy gap, namely the area gap divided by $4G_5$, is
$\cO(N_c^2)$, whereas its flavour correction is $\cO(N_fN_c)$.
The latter therefore does not change the selected branch.  The
horizon-hugging limit of the connected surface produces the usual thermal
volume term \cite{Tonni2011}.  The topology competition relevant to mutual
information is treated separately in section~\ref{sec:mi}.
Evaluating the area variation \eqref{eq:areaVariation} on this surface with
the tensor ansatz \eqref{eq:tensorAnsatz}, and including both radial halves,
gives the difference between the parallel and transverse flavour-entropy
corrections.  Appendix~\ref{app:norm} details the area weights, overall sign
and normalisation.  The result is
\begin{equation}
 \Delta S_{\parallel}-\Delta S_{\perp}
 =-4\pi\Ncal\frac{e^2V_2}{z_h^2}\,\cD(e,\delta;\ell),
 \label{eq:entropyDifference}
\end{equation}
where
\begin{equation}
 K_\ell(u;u_t)\equiv
 \frac{u_t^3}{u^3\sqrt{1-u^4}\sqrt{u_t^6-u^6}}
 \left(1-\frac{u^6}{u_t^6}\right),
 \qquad
 \cD(e,\delta;\ell)=\int_0^{u_t(\ell)}du\,
 K_\ell\bigl(u;u_t(\ell)\bigr)\varphi(u).
 \label{eq:Dfunctional}
\end{equation}
Whenever a strip width appears as the third argument of $\cD$, it is
understood in units of $z_h$.
Since $K_\ell\geq0$ and $-G_T\geq0$, $\cD$ is an integral of the tensor
source with non-negative weight; its zero therefore reflects a balance
between oppositely signed radial contributions.
Our sign convention is therefore
\begin{equation}
 \cD<0\quad\Longleftrightarrow\quad \Delta S_{\parallel}>\Delta S_{\perp}.
\end{equation}
Figure~\ref{fig:source-and-sign} illustrates how the density-dependent radial
sign change of the DBI tensor source is converted by the RT/Green kernel into
a zero of the entropy difference between the two orientations.  We define
the crossover by
\begin{equation}
 \cD(e,\delta_c;\ell)=0.
 \label{eq:deltaCdef}
\end{equation}
Entries labelled $e=0$ in reversal plots and tables denote the smooth $e\to0$ limit of the
normalised order-$e^2$ coefficient after factoring out the overall $e^2$ in
eq.~\eqref{eq:entropyDifference}.  At strictly zero electric field the physical orientation
difference vanishes identically for every density.

\begin{figure}[t]
 \centering
 \subfloat[Crossover density versus strip width at fixed electric field.]{%
  \includegraphics[width=0.48\textwidth]{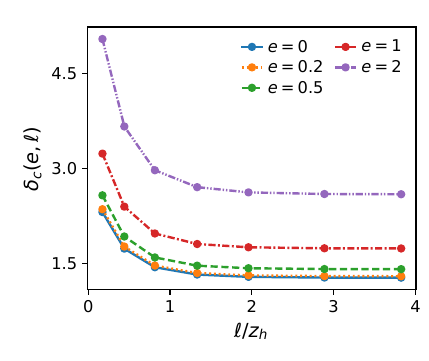}}\hfill
 \subfloat[Interpolated crossover surface with independently computed one-dimensional roots marked explicitly.]{%
  \includegraphics[width=0.48\textwidth]{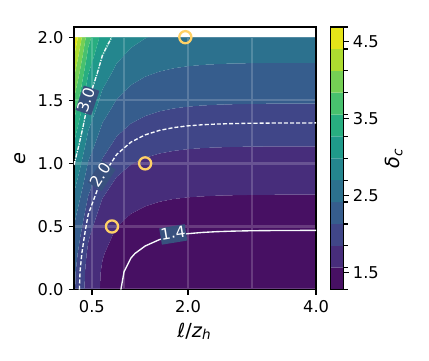}}
 \caption{Density-driven reversal line $\delta_c(e,\ell)$.  With the sign convention of
 eq.~\eqref{eq:entropyDifference}, below the line the parallel strip receives the larger
 flavour-entanglement correction; above it the transverse strip does.  The filled surface and contours
 are visual interpolations, while open symbols mark direct one-dimensional root calculations; the
 quoted digits are fixed by the adaptive quadrature reference described in
 appendix~\ref{app:numerics}.  The curve labelled $e=0$ is the normalised
 smooth $e\to0$ limit; the physical orientation difference vanishes at strictly zero field.  Line styles
 supplement colour in panel (a) and distinguish the labelled contours in panel (b).}
 \label{fig:strip-phase}
\end{figure}

Solving this condition gives the reversal lines shown in
figure~\ref{fig:strip-phase}.  Over the investigated range they exhibit the
monotonic trends
\begin{equation}
 \partial_e\delta_c>0\quad(e>0),\qquad \partial_\ell\delta_c<0.
\end{equation}
Both inequalities describe numerical trends in the investigated range;
we do not assert a global monotonicity theorem.
Both $\delta_c$ and $\delta_{\mathrm{loc}}$ below are even in $e$, so their first derivatives with
respect to $e$ vanish at $e=0$ and the electric-field trend may equivalently be stated as monotonicity in
$|e|$.
Numerical evaluation of the limiting functional at $u_t=1$ gives, at weak field,
\begin{equation}
 \delta_c(e,\infty)=
 1.2794+0.5725\,e^2-0.165\,e^4+\cO(e^6),
 \label{eq:largeWidthExpansion}
\end{equation}
within the displayed weak-field fit.  The quartic coefficient is obtained
from a polynomial fit in $q_{\mathrm{fit}}=e^2$; appendix~\ref{app:numerics} specifies the
fit window, degree and uncertainty estimate.

\subsection{Semi-analytic interpretation}

The sign of the local source in eq.~\eqref{eq:tensorSource} changes at
\begin{equation}
 \delta_{\mathrm{loc}}(u,e)=\frac{s}{u\sqrt{1+s u^2}}.
 \label{eq:deltaLocal}
\end{equation}
It obeys
\begin{equation}
 \partial_u\delta_{\mathrm{loc}}<0,\qquad
 \partial_e\delta_{\mathrm{loc}}>0\quad(e>0),
\end{equation}
which provides qualitative intuition for the observed trends: larger strips
reverse at lower densities, and the reversal occurs at higher densities for
larger electric-field magnitudes.
To characterise the broad radial kernel, define $u_{\mathrm{eff}}(\ell)$ at weak field by
\begin{equation}
 \delta_c(0,\ell)=\frac{1}{u_{\mathrm{eff}}\sqrt{1+u_{\mathrm{eff}}^2}}.
\end{equation}
It approaches $u_{\mathrm{eff}}\simeq0.654$ at large width: although the RT surface approaches the
horizon, the factor $1-u^6/u_t^6$ suppresses the immediate turning-point region in the orientation
difference.  The quantity $u_{\mathrm{eff}}$ is a diagnostic proxy for this broad radial kernel, not a
claim that the response is localised on a sharply defined radial shell.

\begin{figure}[t]
 \centering
 \subfloat[Effective radial shell inferred from the weak-field crossover.]{%
  \includegraphics[width=0.48\textwidth]{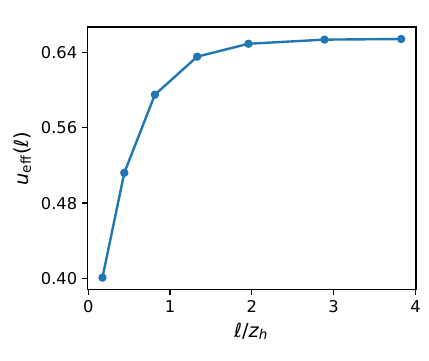}}\hfill
 \subfloat[Exact weak-field slope and local-shell approximation.]{%
  \includegraphics[width=0.48\textwidth]{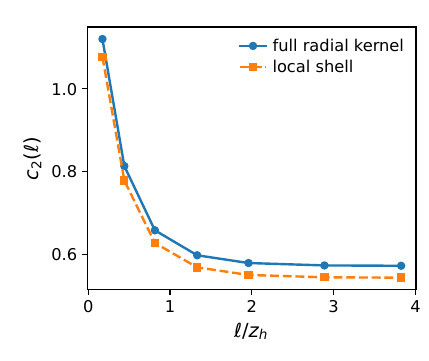}}
 \caption{Semi-analytic description of the crossover, using the sign convention of
 eq.~\eqref{eq:entropyDifference}.  The effective radial coordinate is a diagnostic representation
 of the broad kernel, not a sharply localised bulk position.  The local sign-change estimate captures
 most of the electric-field dependence; the remaining difference reflects the finite width of the
 graviton/RT kernel and the electric-field dependence of the source prefactor.  Both numerical curves use
 fixed-order quadrature of order $N=720$.}
 \label{fig:semi-analytic}
\end{figure}

Writing
\begin{equation}
 \delta_c(e,\ell)=\delta_0(\ell)+c_2(\ell)e^2+\cO(e^4),
\end{equation}
implicit differentiation gives the weak-field slope without a finite-field root scan:
\begin{equation}
 c_2(\ell)=-\left.
 \frac{\partial_{e^2}\cD}{\partial_\delta\cD}
 \right|_{e=0,\delta=\delta_0(\ell)}.
 \label{eq:c2formula}
\end{equation}
For the local-shell approximation, we keep $u_{\mathrm{eff}}(\ell)$,
fixed by the zero-field crossover, unchanged as the electric field varies.
Thus
\begin{equation}
 c_{2,\mathrm{loc}}(\ell)
 =\left.\partial_{e^2}\delta_{\mathrm{loc}}
   \bigl(u_{\mathrm{eff}}(\ell),e\bigr)\right|_{e=0}
 =\delta_0(\ell)\frac{2+u_{\mathrm{eff}}^2}
                         {4(1+u_{\mathrm{eff}}^2)}.
 \label{eq:c2localshell}
\end{equation}
The relative discrepancy $|c_{2,\mathrm{loc}}-c_2|/|c_2|$ is
$3.8$--$5.1\%$ over the widths shown in figure~\ref{fig:semi-analytic}.

\section{Mutual-information anisotropy}\label{sec:mi}

Holographic mutual information (MI) is UV finite and exhibits a connected/disconnected surface transition
at finite temperature \cite{FischlerKunduKundu2013}.  Orientation-dependent holographic MI has also
been studied in anisotropic Q-lattice backgrounds \cite{LiuNiuWu2019}; our use below differs in that
only the leading probe-flavour correction is anisotropic.  UV finiteness follows from the entropy
combination defining MI; taking the parallel-minus-transverse difference additionally cancels the
common D7-sourced scalar/trace response.

Consider two equal strips labelled $A$ and $B$, each of width $\ell$ and separated by $h$, and let
$S(\ell')$ denote the entropy of a strip of generic width $\ell'$.  If $S_0(\ell')$ is the background strip entropy, the
background transition separation $h_0(\ell)$ is defined by
\begin{equation}
 2S_0(\ell)=S_0\bigl(h_0(\ell)\bigr)+S_0\bigl(2\ell+h_0(\ell)\bigr).
 \label{eq:h0definition}
\end{equation}
In the connected RT phase,
\begin{equation}
 I(A:B)=2S(\ell)-S(h)-S(2\ell+h).
\end{equation}
Denote the leading D7-induced orientation corrections by $\Delta I_\parallel$ and
$\Delta I_\perp$.  Their anisotropy is
\begin{equation}
 \Delta I_{\parallel}-\Delta I_{\perp}
 =-4\pi\Ncal\frac{e^2V_2}{z_h^2}\,\cM(e,\delta;\ell,h),
\end{equation}
with
\begin{equation}
 \cM(e,\delta;\ell,h)=2\cD(e,\delta;\ell)-\cD(e,\delta;h)
 -\cD(e,\delta;2\ell+h).
 \label{eq:Mfunctional}
\end{equation}
The MI crossover is $\cM(e,\delta_c^{\mathrm{MI}};\ell,h)=0$.
The $e=0$ entries use the same normalised smooth-limit convention as in
section~\ref{sec:strip}; the physical MI orientation difference vanishes at
strictly zero field.

\begin{figure}[t]
 \centering
 \includegraphics[width=0.95007\textwidth]{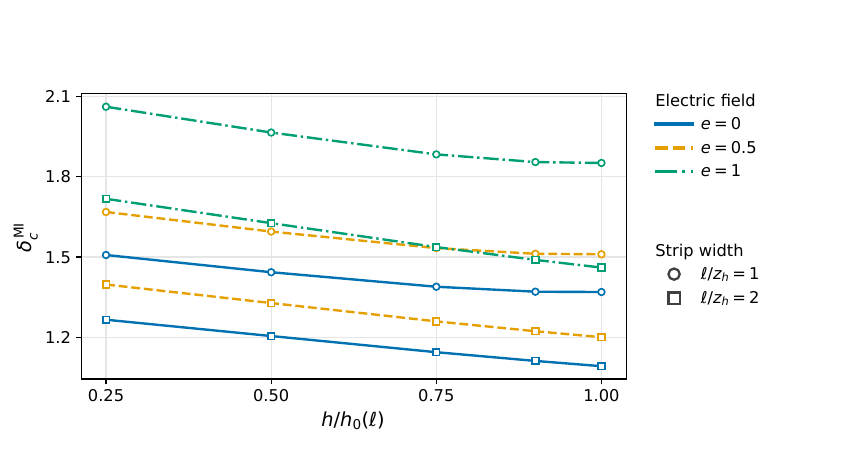}
 \caption{Mutual-information anisotropy reversal as a function of the separation relative to the
 background connected/disconnected transition $h_0(\ell)$.  Mutual information is UV finite; its
 parallel-minus-transverse difference additionally removes the common D7-sourced scalar/trace
 response, providing a complementary UV-finite probe of the strip crossover.  Colour and line style encode the
 electric field, while marker shape encodes the strip width.  The $e=0$ curves are normalised
 smooth-limit data.  Values at $h/h_0=1$ are connected-candidate limits at the
 background transition.  The physical transition is shifted at probe order,
 as described in eq.~\eqref{eq:MIshift}.}
 \label{fig:mi-reversal}
\end{figure}

Figure~\ref{fig:mi-reversal} shows the connected-side crossover as the
separation approaches the background topology transition.  Values plotted
at $h/h_0=1$ are limits of the connected-candidate correction as
$h\to h_0^-$, where the background mutual information becomes nonanalytic.
The physical transition shifts to $h_c=h_0+\Delta h_c$ at probe order;
its shift is treated separately below.

For example, at $\ell/z_h=1$ and $h/z_h=0.2$,
\begin{equation}
 \delta_c^{\mathrm{MI}}=1.4586+0.6585\,e^2+\cO(e^4),
\end{equation}
and the finite-field values are
\begin{equation}
 \bigl(\delta_c^{\mathrm{MI}}(0),\delta_c^{\mathrm{MI}}(0.2),
 \delta_c^{\mathrm{MI}}(0.5),\delta_c^{\mathrm{MI}}(1)\bigr)=
 (1.4586,1.4846,1.6127,1.9884).
\end{equation}
The MI crossover is geometry dependent and need not coincide with the single-strip crossover.

The flavour correction also shifts the connected/disconnected transition.  Let
$\Delta h_c^\parallel$ and $\Delta h_c^\perp$ denote its leading shifts for the two orientations.  The
orientation-dependent shift difference can be written
\begin{equation}
 \frac{\Delta h_c^{\parallel}-\Delta h_c^{\perp}}{z_h}
 =-16\pi G_5\Ncal e^2
 \frac{\cM(e,\delta;\ell,h_0)}{u_t(h_0)^{-3}+u_t(2\ell+h_0)^{-3}}.
 \label{eq:MIshift}
\end{equation}
The expansion about the background topology transition is derived in
appendix~\ref{app:MIshift}.
We regard eq.~\eqref{eq:MIshift} as a secondary observable; the main MI result is the UV-finite
orientation difference within the connected phase.

\section{Numerical validation}\label{sec:numerics}

We implemented the tensor problem in three independent forms: split Green-kernel quadrature,
direct integration of the flux form of eq.~\eqref{eq:tensorODE}, and Chebyshev collocation after
analytic subtraction of the universal UV logarithm.  The principal cross-checks are summarised in
table~\ref{tab:numerical-audit}; appendix~\ref{app:numerics} gives the endpoint maps, kernel split and
convergence details, and figure~\ref{fig:cheb} shows a representative Chebyshev convergence sequence.

\begin{table}[t]
 \centering
 \begin{tabularx}{\textwidth}{@{}Xr@{}}
  \toprule
  Comparison & Absolute discrepancy\\
  \midrule
  Green kernel versus direct flux & $1.3\times10^{-9}$\\
  Green kernel versus Chebyshev & $1.6\times10^{-8}$\\
  Three strip roots versus adaptive reference & $3.2\times10^{-7}$\\
  Weak-field slope versus the $e=0.02$ estimate & $8.9\times10^{-5}$\\
  \bottomrule
 \end{tabularx}
 \caption{Independent numerical comparisons.  The first two rows refer to
 distinct single-point functional evaluations; the last two report maxima
 over the sample sets specified in appendix~\ref{app:numerics}.
 These discrepancies measure
 agreement between methods and do not replace the adopted uncertainty of
 a quoted quantity.}
 \label{tab:numerical-audit}
\end{table}

For the tensor and MI crossover roots covered by this audit, the adopted absolute uncertainties range from $2\times10^{-6}$ to
$5\times10^{-6}$.  The corresponding uncertainties are $3\times10^{-6}$ for the large-width
quadratic coefficient, $2\times10^{-4}$ for the quartic coefficient, $2\times10^{-6}$ for
$u_{\mathrm{eff}}$, and $1\times10^{-5}$ for the MI quadratic coefficient.  The displayed digits
are limited accordingly; solver tolerances are not interpreted as error estimates.
The OSM diagnostics in section~\ref{sec:osm-comparison} and the auxiliary
scalar coefficients in table~\ref{tab:Cmin} are separate evaluations and
are not covered by these uncertainty estimates.

\section{Discussion}\label{sec:discussion}

\subsection{What reverses?}

At fixed $T$, $E$ and $\ell$, $\delta_c$ marks the reversal of the ordering
of the two flavour-entropy corrections.  The steady solution and the tensor
boundary-value problem remain smooth there, so this is a response crossover,
not a thermodynamic phase transition.  It may be viewed heuristically as a
density-tuned competition between charge-neutral-sector and net-density
contributions to the anisotropic stress.  It is not determined by equality
of the two terms in the current formula: the full radial stress profile is
weighted by the closed-string graviton and RT kernels.  The effective-shell
construction captures the qualitative electric-field and strip-width trends,
while the exact reversal line retains this non-local radial information.

\subsection{Relation to the effective temperature}

The open-string horizon and its effective temperature $\Tstar$ govern
probe fluctuations, dissipation and noise
\cite{KimShockTarrio2011,NakamuraOoguri2013,HoshinoNakamura2015}.
The observable computed here instead measures the closed-string metric
response to the D7 stress anisotropy.  At fixed bath temperature and strip
width, the electric field changes the radial profile of the DBI source and
its overall prefactor, while the background Green and strip kernels,
and the background turning point $u_t$, remain independent of the field.
A common scalar contribution, including any isotropic extensive term,
cancels from the orientation difference.  The anisotropy alone therefore
does not test a proposed coefficient for that common term.

The time derivative used to study entanglement rates
\cite{OBannonProbstRodgersUhlemann2017} removes static contributions,
whereas the orientation difference used here removes the common isotropic
response.  The entanglement temperature in the rate relation depends on the
background and subregion, and is distinct from $\Tstar$.

Beyond fluctuation relations, a D3--D7 model with electric and magnetic
fields exhibits Landau susceptibility and correlation-length exponents
along a fixed-$\Tstar/T$ path towards a current-driven tricritical point
\cite{KanazawaMatsumotoNakamura2026}.
For relativistic NESSs, a frame-independent proper effective temperature has
been proposed as a natural scalar characterisation
\cite{HoshinoNakamura2020}.  In a moving D3--D5 defect, order-parameter data
collapse when expressed in terms of this quantity
\cite{NakamuraOkabayashi2025}.  The electrically driven D3--D7 state
considered here is not a boosted equilibrium state, and whether its absolute
scalar entanglement coefficient is controlled by an analogous proper variable
remains open.

At positive bath temperature, equilibrium probe thermodynamics supplies
a thermal entropy density \cite{KarchOBannonThermodynamics2007}, while the
strict zero-temperature finite-density probe calculation has shape-dependent
volume terms \cite{ChangKarchUhlemann2014}.  Appendix~\ref{app:scalar-volume}
illustrates why extending scalar entropy coefficients to the driven state
requires both a conserved source prescription and a state-matching condition.
For its finite-temperature auxiliary source, fixing the reference horizon
and fixing the Hawking temperature give different area responses and can
even change the sign of the electric correction.  These choices are not
fixed by $\Tstar$.  At zero temperature and zero density the driven probe still has
$\Tstar\propto\sqrt E>0$
\cite{NakamuraOoguri2013,HoshinoNakamura2015}.
In appendix~\ref{app:scalar-zero-temperature} we subtract the ingoing null
component responsible for secular Joule heating and add no homogeneous mass
mode.  The resulting stationary first-order correction approaches an
area-like form in the controlled wide-strip regime
$1\ll\sqrt E\,\ell\ll\kappa^{-1}$, where $\kappa=8\pi G_5\Ncal$,
with no extensive term proportional to $\Ncal E^{3/2}V_2\ell$.
Thus the effective thermality of probe fluctuations is not accompanied by a
corresponding extensive term in this specified heating-subtracted response.
This statement concerns the leading probe correction, rather than the
fixed-$\kappa$ asymptotics of a fully backreacted NESS.

\subsection{Relation to open-string areas}\label{sec:osm-comparison}

Related caution about assigning an entropy to the worldvolume horizon
already appeared in ref.~\cite{OBannonProbstRodgersUhlemann2017}.
Ref.~\cite{BanerjeeEtAl2021} explored RT-like area prescriptions in
open-string geometries.  Recent work extends geometric OSM calculations to
mutual information, entanglement wedge cross-sections and negativity in
three- and four-dimensional open-string geometries
\cite{PaulRoyChowdhuryGangopadhyay2026}, interpreting them as flavour-sector
information measures.  We do not adopt this identification for the leading
flavour correction to boundary spatial entanglement entropy: the OSM's role
in probe fluctuation dynamics does not by itself establish an entropy
functional.  The static comparison in ref.~\cite{OkabayashiPaperI} already
found disagreement with controlled flavour entanglement.  Here we test the
driven state with two area diagnostics in the external five-dimensional
open-string metric,
\begin{equation}
 G^{\mathrm{os}}_{ab}=g_{ab}-F_{ac}g^{cd}F_{db},\qquad
 \mathcal A_{\mathrm{os}}[\Sigma]
 =\int_\Sigma d^3\sigma\,
 \sqrt{\det\!\left(G^{\mathrm{os}}_{ab}
              \partial_mX^a\partial_nX^b\right)}.
 \label{eq:osmDiagnosticDefinition}
\end{equation}
Here $g$ is the unperturbed induced external metric, $F$ follows
section~\ref{sec:setup}, and $\sigma^m$ parametrise the three-dimensional
spacelike surface $\Sigma$.  No additional radial Weyl, effective-dilaton
or DBI-determinant weight is included.  The undeformed internal D7 $S^3$
would contribute only the constant volume $2\pi^2$ in units $L=1$, which
is suppressed.  Let $M$ be the $(\bar t,\bar x)$ block of $G^{\mathrm{os}}$
and $\mathbf v=(G^{\mathrm{os}}_{\bar t u},G^{\mathrm{os}}_{\bar x u})^\top$.
We introduce shifted coordinates $\mathsf T,\mathsf X$ by
\begin{equation}
 \binom{\bar t}{\bar x}=\binom{\mathsf T}{\mathsf X}
             +\binom{\chi_t(u)}{\chi_x(u)},\qquad
 \binom{\chi_t'}{\chi_x'}=-M^{-1}\mathbf v,\qquad
 \chi_t(0)=\chi_x(0)=0,
 \label{eq:osmDiagnosticCoordinates}
\end{equation}
where primes denote $u$ derivatives.  These shifts remove both radial
cross terms and identify $\mathsf T,\mathsf X$ with boundary time and
position.  A mixed component remains:
$G^{\mathrm{os}}_{\mathsf T\mathsf X}=-\delta j\mathcal X u^4$ in our convention.

The constant-time prescription fixes $\mathsf T=0$ before extremising the
spatial embedding; the covariant prescription varies the full spacelike
embedding at equal boundary times.  The anchors are
$\mathsf X=\pm\ell/(2z_h)$ for the parallel strip or
$\bar y=\pm\ell/(2z_h)$ for the transverse strip, at $\mathsf T=0$, with
the other two spatial directions extended.  Each recorded surface lies
on the connected exterior branch and has a smooth turning point
$0<u_t^{\mathrm{os}}<u_*$, solved separately for each orientation and
prescription to match the full boundary width, not the closed-string
turning point.  Equal boundary time for the covariant parallel surface
requires $\int_0^{u_t^{\mathrm{os}}}\mathsf T'(u)\,du=0$; the transverse
covariant surface reduces to its constant-$\mathsf T$ counterpart.
The comparison matches $e,\delta,\ell/z_h$ and the transverse coordinate
area.  We use a common cutoff $u=\epsilon$ and subtract the common
$\epsilon^{-2}$ divergence of each area per unit dimensionless transverse
coordinate area before taking the orientation difference.  A common
multiplicative normalisation does not affect its zero; no physical OSM
Newton constant or entropy identification is assumed.  These are distinct
variational prescriptions, not coordinate descriptions of one covariant
functional.

Both show a qualitative density-driven orientation reversal, but their
crossover densities disagree with the controlled result and with each other.
At $(e,\ell/z_h)=(0.5,0.8138)$ they are approximately $1.5988$, $1.8304$ and
$2.1016$ for the controlled, constant-time and covariant prescriptions.
These normalisation-independent mismatches rule out either tested bare OSM
prescription as a formula for the leading flavour correction to boundary
spatial entanglement entropy in this D3--D7 setting.  This does not exclude a different open-string functional derived from a
boundary replica construction.

\subsection{Relation to other current-carrying NESSs}

The coherent-conductor examples suggest a useful organising picture for
non-thermal extensive subsystem entropy and quantum correlations:
non-equilibrium occupation imbalance allows scattering to partition incoming
modes coherently across the chosen spatial bipartition.  This does not require
inelastic energy exchange.  For the pure global zero-temperature scattering
state in ref.~\cite{FraenkelGoldstein2021}, the volume coefficient of an
interval in one lead is an integral of the binary entropy of the transmission
probability over the bias window; it vanishes for perfect transmission or
perfect reflection throughout that window.  At finite bias, a single impurity
with partial transmission over a finite part of the window is sufficient,
and multiple scattering is not necessary in this geometry.  Contact-induced
occupation mixing \cite{Ribeiro2017} and the far-from-equilibrium,
multiple-scattering enhancement inside a disordered wire
\cite{HakoshimaShimizu2019} provide related, but distinct, realisations.
Equal mirrored intervals on opposite sides of a scatterer also support
extensive fermionic negativity and positive coherent information in the
large-interval regime \cite{FraenkelGoldstein2023}; mutual information
alone does not isolate purely quantum correlations.
A complementary mixed-state example is provided by a Lindblad-driven
noninteracting fermion chain, where a local impurity can induce nonzero
steady-state fermionic negativity between separated end segments
\cite{NavaEtAl2026}.

The survival and spatial support of the resulting coherence are as important
as its generation.  Noninteracting diffusive circuits can retain extensive
quantum correlations, while chaotic interactions in the models studied in
ref.~\cite{GullansHuse2019} favour local equilibration and short-range
entanglement.  In the three-dimensional Anderson problem, mutual coherence
remains extensive at criticality but becomes area-like after localisation
\cite{GullansHuseLocalization2019}.  These results do not imply that more
scattering always increases entanglement.  Nor should the loss of long-range
quantum correlations be confused with the absence of a thermal extensive
term in a mixed-state subsystem entropy.

If the area-like behaviour of appendix~\ref{app:scalar-zero-temperature}
is taken as indicative of the stationary response after secular heating is
removed, one possible interpretation is that the homogeneous setup lacks
explicit scattering centres through which the coherent mode partitioning
of the wire examples is realised
\cite{Ribeiro2017,FraenkelGoldstein2021,FraenkelGoldstein2023,HakoshimaShimizu2019}.
In this picture, spatially resolved disorder could generate an additional
non-thermal volume-law or quasi-volume-law contribution.  This is a heuristic
interpretation, not a derivation of the boundary correlation structure; it
concerns a particular scattering mechanism rather than a general obstruction
to volume terms in homogeneous systems.  Indeed, the homogeneous equilibrium
D3--D7 state at zero temperature and finite density already exhibits
shape-dependent volume terms at leading probe order, without an applied
electric field or explicit scattering centres
\cite{ChangKarchUhlemann2014}.
Momentum relaxation and spatially resolved worldvolume disorder offer
complementary extensions, and comparing entropy scaling with and without
disorder under a common stationary-completion prescription would test this
picture.

\subsection{Scope and limitations}

The main results concern a massless, spatially homogeneous probe at leading
order in $N_f/N_c$ on the finite-temperature AdS$_5$--Schwarzschild background.
They isolate the D7-sourced tensor response; absolute scalar-sector volume
coefficients additionally require a specified stationary completion once
closed-string backreaction is included.
Other open problems include massive embeddings, magnetic fields, the low-temperature finite-density
scaling limit, a fully controlled Lorentzian probe-replica derivation and a direct study of
disorder-driven long-range entanglement.

\section{Conclusion}\label{sec:conclusion}

We have computed a controlled leading probe-flavour entanglement anisotropy in the electric-field-driven
D3--D7 steady state without constructing the full backreacted geometry.  The Chang--Karch graviton
Green-function representation reduces the problem to the full DBI stress anisotropy in a single
closed-string tensor channel.  The orientation-dependent correction reverses sign across a density-driven
line $\delta_c(e,\ell)$ that, over the investigated parameter range, moves to
higher densities with increasing electric-field magnitude and to lower densities
with increasing strip width.  A
mutual-information analogue provides a UV-finite confirmation.  The crossover follows from the radial
sign structure of the DBI tensor source and is confirmed by independent numerical implementations.
These results identify a non-trivial but smooth entanglement-response
crossover in a homogeneous dissipative NESS and establish a controlled starting point for extensions
with mass, magnetic fields and momentum relaxation.

\acknowledgments
The author thanks Hideaki Hakoshima, Masataka Matsumoto and Shin Nakamura
for helpful advice. OpenAI's GPT-5.5, GPT-5.6 Sol and GPT-6 Astra, together
with Anthropic's Claude Fable 5 and 5.1, were used for language editing,
code-related assistance, and exploratory analytical and numerical checks.
All AI-assisted material was independently verified by the author, who
takes full responsibility for the content of this work.

\appendix
\section{Derivation of the full DBI tensor source}\label{app:source}

Set $z_h=L=2\pi\alpha'=1$ in this appendix.  With $M_{ab}\equiv g_{ab}+F_{ab}$ and primes
denoting derivatives with respect to $u$, the $(t,x,u)$ block is
\begin{equation}
 M_{(txu)}=
 \begin{pmatrix}
 -f/u^2 & -e & -A_t'\\
 e & 1/u^2 & -a_x'\\
 A_t' & a_x' & 1/(u^2f)
 \end{pmatrix}.
\end{equation}
Let $M_{5d}$ denote the full external worldvolume block obtained by adjoining
the $y$ and $w$ directions.  Factoring out the determinant of the induced
background metric, define
\begin{equation}
 -\det M_{(txu)}=u^{-6}\mathcal X_{\mathrm{raw}},
 \qquad
 -\det M_{5d}=u^{-10}\mathcal X_{\mathrm{raw}}.
\end{equation}
Thus $\mathcal X_{\mathrm{raw}}$ is the dimensionless determinant ratio
entering the DBI square root, not the complete worldvolume determinant.
The DBI radial factor is
\begin{equation}
 \mathcal X_{\mathrm{raw}}=1-\frac{e^2u^4}{f}-u^4(A_t')^2+u^4f(a_x')^2.
\end{equation}
The first integrals are
\begin{equation}
 A_t'=-\delta u\sqrt{\mathcal X},\qquad
 a_x'=\frac{ju}{f}\sqrt{\mathcal X}.
\end{equation}
The ensemble choice relevant to a replica variation is separate from these
physical first integrals.\footnote{The physical solutions at replica index $n=1$ are
parametrised here by the charge density $d$.  In a genuine probe-replica
variational problem one must separately specify whether the boundary chemical
potential or the charge density $d$ is held fixed as $n$ varies; the
fixed-density choice requires the corresponding Legendre boundary term.
This distinction is not used in the present calculation.}
These first integrals imply
\begin{equation}
 \mathcal X=\frac{f-e^2u^4}{f(1+\delta^2u^6)-j^2u^6}.
\end{equation}
Simultaneous vanishing at $u=u_*$ gives the current quoted in the text and reduces the ratio to
 eq.~\eqref{eq:Xregular}.

Following the probe stress-tensor construction of ref.~\cite{KarchOBannonThompson2009}, the effective
five-dimensional mixed D7 stress tensor after internal-space integration is
\begin{equation}
 T^a{}_b=-\Ncal\sqrt{\mathcal X}\,g_{bc}(M^{-1})^{(ac)}.
\end{equation}
Parentheses on $(ac)$ denote the symmetric part of the inverse DBI matrix.
Because $F$ has no $y$ component,
\begin{equation}
 \cT_T=-\Ncal\sqrt{\mathcal X}\left[g_{xx}(M^{-1})^{xx}-1\right].
\end{equation}
Using the cofactor of $M_{(txu)}$ and the first integrals gives
\begin{equation}
 \cT_T=\Ncal\frac{u^4}{f\sqrt{\mathcal X}}
 \left[j^2u^2\mathcal X-e^2\right],
\end{equation}
which reduces algebraically to eq.~\eqref{eq:tensorSource}.

\section{Normalisation of the strip orientation difference}\label{app:norm}

In dimensionless coordinates $\bar x^\mu=x^\mu/z_h$ and $u=z/z_h$, the D7 action retains the
overall coefficient $\Ncal$.  The physical tensor perturbation and the normalised solution are related
by
\begin{equation}
 \Phi=\frac{32\pi G_5\Ncal e^2}{3}\varphi.
\end{equation}
For one radial half of the unperturbed U-shaped strip surface, let $i=x$ for the parallel strip and
$i=y$ for the transverse strip, and define
\begin{equation}
 \rho(u)\equiv\gamma_{(0)}^{uu}g_{ii}\bigl(\partial_u\bar x^i\bigr)^2
 =\frac{u^6}{u_t^6},\qquad
 K_0(u;u_t)\equiv
 \frac{u_t^3}{u^3\sqrt{1-u^4}\sqrt{u_t^6-u^6}}.
\end{equation}
The main-text kernel in eq.~\eqref{eq:Dfunctional} is therefore
$K_\ell(u;u_t)=K_0(u;u_t)[1-\rho(u)]$.
Using the tensor ansatz in eq.~\eqref{eq:tensorAnsatz}, the induced-area weights on that half are
\begin{align}
 \left.\frac12\gamma_{(0)}^{ab}\delta\gamma_{ab}\right|_{\parallel}
 &=\frac12(\rho-1)\Phi,&
 \left.\frac12\gamma_{(0)}^{ab}\delta\gamma_{ab}\right|_{\perp}
 &=\frac14(1-\rho)\Phi,\\
 \left.\frac12\gamma_{(0)}^{ab}\delta\gamma_{ab}\right|_{\parallel-\perp}
 &=-\frac34(1-\rho)\Phi.&&
\end{align}
Thus the two individual half-surface variations and their difference are
\begin{align}
 \delta A_{\parallel}^{\mathrm{half}}
 &=\frac{V_2}{z_h^2}\int_0^{u_t}du\,K_0\,\frac{\rho-1}{2}\Phi,&
 \delta A_{\perp}^{\mathrm{half}}
 &=\frac{V_2}{z_h^2}\int_0^{u_t}du\,K_0\,\frac{1-\rho}{4}\Phi,\\
 \delta A_{\parallel}^{\mathrm{half}}-\delta A_{\perp}^{\mathrm{half}}
 &=-\frac{3V_2}{4z_h^2}\int_0^{u_t}du\,K_\ell\Phi.&&
\end{align}
The complete U-shaped RT surface has two geometric radial halves.  Adding them and using
$S=A/(4G_5)$ gives the linear area difference
\begin{equation}
 \Delta S_{\parallel}-\Delta S_{\perp}
 =-\frac{3V_2}{8G_5z_h^2}\int_0^{u_t}du\,K_\ell(u;u_t)\Phi(u),
\end{equation}
including both the sign and the factor of two.  Substituting the relation between $\Phi$ and
$\varphi$, the numerical factors assemble as
\begin{equation}
 2\left(-\frac34\right)\frac{1}{4G_5}
 \left(\frac{32\pi G_5\Ncal e^2}{3}\right)
 =-4\pi\Ncal e^2.
\end{equation}
The factors of $G_5$ therefore cancel, and $K_\ell\varphi$ is precisely the integrand defining
$\cD$ in eq.~\eqref{eq:Dfunctional}.  This reproduces eq.~\eqref{eq:entropyDifference} and fixes the
sign convention $\cD<0\Longleftrightarrow\Delta S_{\parallel}>\Delta S_{\perp}$.

\section{Mutual-information transition shift}\label{app:MIshift}

In this appendix all width arguments are dimensionless, in units of $z_h$.
For a generic dimensionless width $\ell'$, define the rescaled background
area density $a_0(\ell')=z_h^2 A_0(z_h\ell')/V_2$, where $A_0$ is the
complete strip area before division by $4G_5$.  A common UV regulator is
understood.  We define
$M_0(\ell,h)=2a_0(\ell)-a_0(h)-a_0(2\ell+h)$ as the background
disconnected-minus-connected candidate area difference.  At the background
transition $h=h_0(\ell)$, $M_0=0$ by
eq.~\eqref{eq:h0definition}.  The Hamilton--Jacobi relation
$da_0/d\ell'=u_t(\ell')^{-3}$ gives
\begin{equation}
 \partial_hM_0=-u_t(h)^{-3}-u_t(2\ell+h)^{-3}.
\end{equation}
Expanding the transition condition to first order in the probe correction gives eq.~\eqref{eq:MIshift}.

\section{Numerical methods and convergence}\label{app:numerics}

This appendix records implementation details and convergence checks for
readers interested in reproducing the calculations or examining the quoted
numerical uncertainties.  The main validation results are summarised in
section~\ref{sec:numerics}; the technical details below are not needed to
follow the physical discussion.

The strip functional is evaluated both with fixed-order Gauss--Legendre quadrature and with an
independent arbitrary-precision adaptive reference.  For the latter we transform the turning-point
coordinate according to $(u/u_t)^3=\sin\theta$.  The functional uses
$\theta=(\pi/2)\zeta^6$, with $\zeta\in[0,1]$, which both removes the apparent square-root singularity and clusters points near
the UV endpoint; the strip-width integral uses the analogous map with
$\theta=(\pi/2)\zeta^3$.  The fixed-order sequence is
$N=180,240,360,480,720$.

For the Green representation, the integration over $u'$ is split explicitly at $u'=u$, where
$\min(u,u')$ changes branch.  Since
$\widehat{\cT}_T(u')/u'^5=-1/u'+\cO(u')$ near the boundary, the universal $-1/u'$ term is integrated
analytically in terms of logarithms and a dilogarithm; only the regular remainder is sent to adaptive
quadrature.  This gives the reference values used to determine the displayed digits.

For the large-width limit we directly evaluate the limiting functional at
$u_t=1$, whose strip weight is
$K_\infty(u)=\sqrt{1-u^6}/[u^3\sqrt{1-u^4}]$ and remains finite as $u\to1$.
This is the limit of the finite-width problem, not a finite-width surface
placed on the horizon; no extrapolation of a sequence of turning points is
used.  The weak-field fit described below is a separate fit in $e^2$.

The adaptive reference uses 30-decimal-digit arithmetic; where implemented,
we compare with a repeat at 25 digits.  These are working precisions, not
numbers of correct digits in the result.  For directly repeated quantities,
we retain the larger of the observed precision-repeat variation and a
prespecified $10^{-11}$ safeguard; the corresponding variation for
$u_{\mathrm{eff}}$ is propagated from the zero-field crossover through its
defining relation.
Separate conservative lower bounds on adopted uncertainties are
$2\times10^{-4}$ for the large-width quartic coefficient and $10^{-5}$
for the MI quadratic coefficient.  These bounds are reporting safeguards,
not measured physical errors.

For the quartic coefficient in eq.~\eqref{eq:largeWidthExpansion}, we set $q_{\mathrm{fit}}=e^2$ and use the seven
abscissae
\begin{equation}
 q_{\mathrm{fit},k}=\frac{k}{150},\qquad k=0,1,\ldots,6
 \quad\left(0\leq q_{\mathrm{fit}}\leq0.04,\quad 0\leq e\leq0.2\right).
\end{equation}
The primary fit is an unweighted least-squares polynomial of degree four in $q_{\mathrm{fit}}$ (through $e^8$),
with the $q_{\mathrm{fit}}^2=e^4$ coefficient reported in eq.~\eqref{eq:largeWidthExpansion}.  Fit-window dependence
is measured by dropping the $q_{\mathrm{fit}}=0.04$ point and repeating the degree-four fit; fit-order dependence is
measured with a degree-five fit to all seven points.  These variations are evaluated for both the
adaptive-reference and Chebyshev sequences.  For this quartic coefficient,
the precision-repeat contribution is assigned the $10^{-11}$ safeguard;
the complete fit is not independently repeated at 25 and 30 digits.
We take the maximum of this safeguard, the fit variations, and the
discrepancies of the $N=720$ and Chebyshev results from the adaptive
reference, then round upward to one significant digit.  The adopted
$2\times10^{-4}$ uncertainty is the larger of that rounded maximum and
the prespecified lower bound.

As a first independent check, integrating eq.~\eqref{eq:tensorODE} once gives
\begin{equation}
 \frac{1-u^4}{u^3}\varphi'(u)=\int_u^1dv\,\frac{\widehat{\cT}_T(v)}{v^5},
\end{equation}
which is integrated independently with $\varphi(0)=0$.

\begin{figure}[t]
 \centering
 \includegraphics[width=0.62\textwidth]{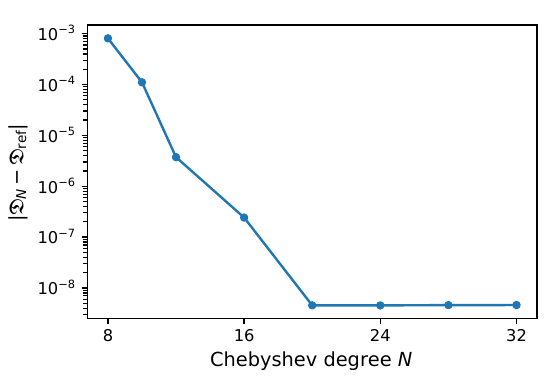}
 \caption{Chebyshev convergence of the strip functional at
 $(e,\delta,\ell/z_h)=(0.5,1,1)$.  Subtracting the universal $u^4\log u$ term yields rapid
 convergence to the split adaptive Green-kernel reference.
 Here $\cD_N$ is evaluated at Chebyshev degree $N$, and
 $\cD_{\mathrm{ref}}$ is the adaptive Green-kernel reference.
 The outer quadrature order is fixed at $720$.}
 \label{fig:cheb}
\end{figure}

As a second check, the UV behaviour $\widehat{\cT}_T=-u^4+\cO(u^6)$ produces
$\varphi=(u^4/4)\log u+\cO(u^4)$.  We subtract the analytic singular part
\begin{equation}
 \varphi_{\log}(u)=\frac1{16}\left[\operatorname{Li}_2(1-u^4)-\frac{\pi^2}{6}\right],
\end{equation}
where $\operatorname{Li}_2$ is the dilogarithm.  We then solve for the analytic remainder on a
Chebyshev--Lobatto grid.  The convergence of the resulting strip functional
to the adaptive Green-kernel reference is shown in
figure~\ref{fig:cheb}.

Across the tensor/MI quantities and weak-field fit coefficients included in this audit, the largest absolute difference between the $N=720$ result and the
adaptive reference is $4.43\times10^{-6}$, while the corresponding maximum Chebyshev difference is
$5.67\times10^{-6}$.  The first row of table~\ref{tab:numerical-audit}
compares the Green and direct-flux evaluations of $\cD$ at
$(e,\delta,\ell/z_h)=(0.5,1,0.8138041)$, with absolute difference
$1.3\times10^{-9}$.  The second compares Green and degree-32 Chebyshev
evaluations at $(1,2,1.3309875)$, with difference $1.6\times10^{-8}$.
These are distinct single-point checks, not maxima over a parameter region.

The third row reports the maximum discrepancy from the adaptive reference
over both the $N=720$ and Chebyshev roots at the parameter pairs
$(e,\ell/z_h)=(0.5,0.8138041)$, $(1,1.3309875)$ and $(2,1.9586629)$,
the three direct-root symbols in figure~\ref{fig:strip-phase}(b).
All lie within $3.2\times10^{-7}$ of the reference.
The fourth row compares the weak-field slope in eq.~\eqref{eq:c2formula}
with the $e=0.02$ finite-field estimate at
$\ell/z_h=0.4402893,0.8138041,1.3309875,1.9586629$.
The maximum discrepancy is $8.9\times10^{-5}$; it includes finite-field
truncation and is not solely a solver error.
Adopted uncertainties conservatively include the measured method discrepancy
and, for fitted coefficients, the fit-window dependence; quadrature and root-solver tolerances are not
reported as physical precision.

\section{Auxiliary scalar response and temperature matching}
\label{app:scalar-volume}

This appendix provides supplementary calculations motivated by the area-
and volume-law behaviour discussed in earlier studies of NESS entanglement
(see section~\ref{sec:discussion}).  None of the tensor-anisotropy or
mutual-information results in sections~\ref{sec:probeEE}--\ref{sec:mi}
relies on these calculations; readers concerned only with those results may
omit this appendix.  Subsections~\ref{app:scalar-source}--\ref{app:scalar-coefficients}
specify a conserved five-dimensional scalar source and compare its
reference-horizon area response with the result at fixed Hawking temperature.
Subsection~\ref{app:scalar-zero-temperature} considers a separate
zero-temperature, zero-density null-subtracted problem.  These are algebraic
source prescriptions, not dynamical reservoir models or ten-dimensional
scalar uplifts.

At $e=0$, differentiating the canonical probe free energy at fixed charge
density $d$ gives the equilibrium entropy density
\cite{KarchOBannonThermodynamics2007},
\begin{equation}
 s_{\mathrm{flavour}}^{\mathrm{eq}}(d,T)
 =\pi\Ncal z_h^{-3}\sqrt{1+\delta^2}.
\end{equation}
At positive bath temperature, this thermodynamic density supplies the
asymptotic large-region extensive coefficient, after temperature matching
\cite{ChangKarchUhlemann2014}.  This thermal limit is distinct from the
shape-dependent finite-density volume terms in the strict zero-temperature
leading-probe calculation.
An electric correction must additionally specify the source completion and
the state-matching condition.  We compute the order-$e^2$ correction below
at fixed $d$ and reference $z_h$, choosing the homogeneous mode so that the
physical Hawking temperature remains $T_0=(\pi z_h)^{-1}$.

\subsection{EF coordinates and conserved source}
\label{app:scalar-source}
Use dimensionless ingoing EF coordinates $(v,\bar x,\bar y,\bar w,u)$ and
the scalar metric
\begin{equation}
 ds^2=\frac{1}{u^2}\left[
 -\{f(u)+a(u)\}\,dv^2-2\,dv\,du+
 \{1+Q(u)\}\,d\bar{\boldsymbol x}^{\,2}\right],
 \qquad f(u)=1-u^4 .
 \label{eq:scalarEFmetric}
\end{equation}
The blackening perturbation $a$ and the spatial perturbation $Q$ are both
of order $\kappa_e=8\pi G_5\Ncal e^2$.  The EF cross term is fixed as a
radial gauge choice, and the boundary time and spatial metric are normalised
by $a(0)=Q(0)=0$.  We also fix the residual radial shift by $Q'(0)=0$.
Primes in subsections~\ref{app:scalar-source}--\ref{app:scalar-coefficients}
denote $u$ derivatives.

Taking the trace with the unperturbed five-dimensional metric, define
\begin{equation}
 \widetilde T_{MN}=T_{MN}-\frac13g^{(0)}_{MN}T^P{}_P .
\end{equation}
Let $\mathcal S_{MN}^{(2)}$ denote the stationary isotropic
trace-reversed source in the electric correction,
$\delta\widetilde T_{MN}^{\mathrm{aux}}=\Ncal e^2\mathcal S_{MN}^{(2)}$.
Its components are specified by
\begin{equation}
 \mathcal S_{uu}^{(2)}=r,\qquad
 \mathcal S_{ii}^{(2)}=c=0,\qquad
 \mathcal S_{vu}^{(2)}=b,\qquad
 \mathcal S_{vv}^{(2)}=fb,\qquad
 b=\left(1+\frac{u^4}{3}\right)r+\frac{u^5-u}{6}r',
 \label{eq:scalarCompletion}
\end{equation}
with $i\in\{\bar x,\bar y,\bar w\}$ and no sum on $i$.
This choice gives a conserved ordinary stress tensor after undoing trace
reversal.  More generally its Bianchi relation is
$b=(1+u^4/3)r+(u^5-u)r'/6+uc'/2$; conservation does not fix $c$.
Equation~\eqref{eq:scalarCompletion} makes the specific choice $c=0$ and
has vanishing radial energy flux.

For $H(u)=1+\delta^2u^6$ and $C_\delta=\sqrt{1+\delta^2}$, we retain the
D7 radial scalar source
\begin{align}
 r(u,\delta)=\mathcal S_{uu}^{(2)}(u;\delta)
 &=\frac{u^2[C_\delta u-\sqrt{H(u)}]^2}
 {(1-u^4)^2\sqrt{H(u)}} \notag\\
 &=\frac{u^2[\delta^2u^2(1+u^2)-1]^2}
 {(1+u^2)^2\sqrt{H(u)}[C_\delta u+\sqrt{H(u)}]^2}.
 \label{eq:scalarSourceMin}
\end{align}
The second form removes the apparent horizon pole and makes nonnegativity
manifest.  In particular,
\begin{equation}
 \lim_{u\to1}\mathcal S_{uu}^{(2)}(u;\delta)
 =\frac{(2\delta^2-1)^2}{16(1+\delta^2)^{3/2}}.
 \label{eq:scalarSourceHorizon}
\end{equation}
Both $r$ and the choice $c=0$ can be checked directly from the DBI expansion.
Writing $q=\sqrt{\mathcal X}$, its EF radial component is
$\widetilde T_{uu}/\Ncal=u^2(e-juq)^2/(f^2q)$.
At fixed density $j=eC_\delta+\cO(e^3)$ and
$q=H^{-1/2}+\cO(e^2)$, which gives
eq.~\eqref{eq:scalarSourceMin}.  The exact covariant spatial scalar average is
\begin{equation}
 \frac{\widetilde T_{\bar x\bar x}
       +\widetilde T_{\bar y\bar y}
       +\widetilde T_{\bar w\bar w}}{3\Ncal}
 =\frac{q^{-1}+Hq}{3u^2}.
 \label{eq:scalarSpatialAverage}
\end{equation}
Its first derivative with respect to $q$ vanishes at $q=H^{-1/2}$, so
its order-$e^2$ coefficient is zero for every fixed $\delta$.
A single transverse component is not this scalar average.
The remaining components in eq.~\eqref{eq:scalarCompletion} specify the
external stationary completion; they are not an assertion that the driven
D7 source alone has no energy flux.

\subsection{Field equations and state matching}
\label{app:scalar-matching}
In the gauge \eqref{eq:scalarEFmetric},
$\delta(R_{uu}+4g_{uu})=-3Q''/2$.  Thus the radial equation is
\begin{equation}
 Q''(u)=-\frac23\kappa_e\,\mathcal S_{uu}^{(2)}(u;\delta),
 \qquad Q(0)=Q'(0)=0.
 \label{eq:scalarQfirstpass}
\end{equation}
The spatial Einstein equation supplies the missing blackening response:
\begin{equation}
 ua'-4a=-\frac{\kappa_e}{3}u^2fr-(3u-u^5)Q'.
 \label{eq:scalarBlackening}
\end{equation}
Together with the Bianchi relation, these equations also satisfy the $vu$
and $vv$ equations.  For example, the solution for $a$ is
\begin{equation}
 a(u)=u^4\left[a(1)+\int_1^u d\hat u\,
 \left\{-\frac{\kappa_e f(\hat u)r(\hat u)}{3\hat u^3}
 -\left(\frac{3}{\hat u^4}-1\right)Q'(\hat u)\right\}\right].
 \label{eq:scalarMassSolution}
\end{equation}
It is regular at the reference horizon and vanishes at the boundary;
the finite coefficient of the homogeneous $u^4$ mass mode remains to be
fixed.  The logarithmic boundary term is of order $u^4\log u$ and does
not change the boundary lapse.

Let the perturbed horizon be $u_H=1+\eta_H$.  Boundary-normalised
surface gravity and horizon area give, to first order,
\begin{equation}
 \eta_H=\frac{a(1)}4,\qquad
 \frac{\delta T_H}{T_0}=\frac{3a(1)-a'(1)}4,\qquad
 \Delta s^{\mathrm{aux}}
 =\frac{\frac32Q(1)-3\eta_H}{4G_5z_h^3}.
 \label{eq:scalarHorizonMatching}
\end{equation}
Temperature matching is essential when comparing a large-strip area with
an entropy density; see eqs.~(50)--(51) of
ref.~\cite{ChangKarchUhlemann2014}.  Here
eq.~\eqref{eq:scalarBlackening} gives $a'(1)=4a(1)-2Q'(1)$.
Fixing the reference horizon by $a(1)=0$ therefore changes the Hawking
temperature by $\delta T_H/T_0=Q'(1)/2$.
Fixing the physical temperature instead selects $a(1)=2Q'(1)$ and
$\eta_H=Q'(1)/2$.

\subsection{The radial diagnostic and the fixed-temperature coefficient}
\label{app:scalar-coefficients}
Integrating eq.~\eqref{eq:scalarQfirstpass} twice gives
\begin{equation}
 Q(1)=\kappa_e C_Q(\delta),
 \qquad
 C_Q(\delta)=-\frac23\int_0^1du\,(1-u)\mathcal S_{uu}^{(2)}(u;\delta).
 \label{eq:CminIntegral}
\end{equation}
The coefficient $C_Q=Q(1)/\kappa_e$ measures the dimensionless spatial-metric
response at the reference horizon.  The area element of its three spatial
directions changes as $(1+Q)^{3/2}=1+3Q/2+\cO(Q^2)$; combined with
$\kappa_e=8\pi G_5\Ncal e^2$, this gives the factor of three in the
reference-horizon entropy diagnostic,
\begin{equation}
 \Delta s_{\mathrm{ref}}^{\mathrm{aux}}
 =\frac{1}{4G_5z_h^3}\frac32Q(1)
 =3\pi\Ncal z_h^{-3}e^2C_Q(\delta).
 \label{eq:scalarHorizonArea}
\end{equation}
This is not the fixed-temperature electric entropy correction.
For the fully specified $c=0$ auxiliary source, temperature matching gives
\begin{equation}
 \Delta s_T^{\mathrm{aux}}
 =\pi\Ncal z_h^{-3}e^2 C_T(\delta),\qquad
 C_T(\delta)=2\int_0^1du\,u\,r(u,\delta).
 \label{eq:scalarVolumeCoefficient}
\end{equation}
All quantities in these two equations refer to the retained order $e^2$.
At zero density the distinct coefficients are
\begin{equation}
 C_Q(0)=\frac{\pi+\log2-4}{12}<0,
 \qquad
 C_T(0)=\frac{\pi+2\log2-4}{8}>0.
 \label{eq:scalarZeroDensity}
\end{equation}
Thus $3C_Q(0)=(\pi+\log2-4)/4$ is negative, whereas the
fixed-temperature correction for this same source is positive.
This is a change of comparison condition, not a change in the integral
$C_Q$.  A different conserved spatial scalar completion would give
$C_T=2\int_0^1du\,u r+\frac32c(1)$, so neither sign is a universal
D3--D7 NESS entropy prediction.

For fixed $0<u<1$, the source has the outer large-density expansion
\[
 \mathcal S_{uu}^{(2)}(u;\delta)
 =\delta\,\frac{u}{(1+u^2)^2}+o(\delta)
 \qquad(\delta\to\infty,\;u\ \text{fixed}).
\]
The $u=\cO(\delta^{-1/3})$ ultraviolet boundary layer contributes at most
$\cO(\delta^{1/3})$ to the integral in eq.~\eqref{eq:CminIntegral}, and is therefore
subleading to the $\cO(\delta)$ outer contribution.
Inserting the fixed-$u$ expansion into
eq.~\eqref{eq:CminIntegral} and using
\[
 \int_0^1du\,\frac{u(1-u)}{(1+u^2)^2}=\frac{4-\pi}{8}
\]
gives
\begin{equation}
 C_Q(\delta)=-\frac{4-\pi}{12}\,\delta+o(\delta),
 \qquad
 \frac{3C_Q(\delta)}{\sqrt{1+\delta^2}}
 \longrightarrow-\frac{4-\pi}{4}.
 \label{eq:CminLargeDensity}
\end{equation}
The fixed-temperature coefficient for the declared $c=0$ source has a
separate large-density limit.  For $\delta\geq1$ the regular expression
for $r$ gives $0\leq 2ur/\delta\leq4u^2+4u$, so its pointwise limit
can be integrated by dominated convergence.  Consequently,
\begin{equation}
 C_T(\delta)=\frac{\pi-2}{4}\,\delta+o(\delta),
 \qquad
 \frac{C_T(\delta)}{\sqrt{1+\delta^2}}
 \longrightarrow\frac{\pi-2}{4},
 \label{eq:CTLargeDensity}
\end{equation}
since $2\int_0^1du\,u^2/(1+u^2)^2=(\pi-2)/4$.
The latter ratio gives the coefficient of $e^2$ in the fractional correction
$\Delta s_T^{\mathrm{aux}}/s_{\mathrm{flavour}}^{\mathrm{eq}}$.
These are weak-field asymptotics of the specified auxiliary source, valid
while the probe response remains small; they do not describe the
fixed-$E_{\mathrm{phys}}$ zero-temperature limit.

Representative values of both coefficients are listed in table~\ref{tab:Cmin}.
\begin{table}[t]
 \centering
 \begin{tabular}{@{}rrr@{}}
  \toprule
  $\delta$ & $C_Q(\delta)$ & $C_T(\delta)$\\
  \midrule
  $0$ & $-0.0137717$ & $0.0659859$\\
  $1$ & $-0.0056024$ & $0.0132446$\\
  $2$ & $-0.0080050$ & $0.1114730$\\
  $5$ & $-0.0901510$ & $0.8227196$\\
  \bottomrule
 \end{tabular}
 \caption{Auxiliary scalar-response coefficients for the specified $c=0$
 source.  The spatial-metric coefficient $C_Q=Q(1)/\kappa_e$ gives the
 reference-horizon area diagnostic with normalised coefficient $3C_Q$;
 $C_T$ instead includes the horizon shift required by fixed-temperature
 matching.  Neither comparison defines a completion-independent NESS
 entropy coefficient.}
 \label{tab:Cmin}
\end{table}

\subsection{Zero temperature at zero density}
\label{app:scalar-zero-temperature}
At $d=T=0$, keep the rescaled electric field $E>0$ fixed and use
$\xi=\sqrt E\,z$.  This is not the zero-temperature limit of the weak-$e$
expansion above.  The unthermostatted probe transfers energy at the Joule rate
$EJ$ \cite{KarchOBannonThompson2009}; the entanglement-rate calculation in
ref.~\cite{OBannonProbstRodgersUhlemann2017} is a direct precedent for its
heating response.  To separate a stationary electric response from secular
Joule heating, we define the source subtraction
$\Delta T=T_{\mathrm{D7}}(E)-T_{\mathrm{D7}}(0)-T^{\mathrm{null}}$, where
$T^{\mathrm{null}}_{MN}dX^M dX^N=\Ncal E^{5/2}z^3\,dv^2$
in units $L=1$, with $v=t-z$.  The null component generates a Vaidya-type
heating response.  Its subtraction leaves the conserved stationary source
used below.

The exact massless DBI solution has $J=\Ncal E^{3/2}$ and
$q(\xi)=\sqrt{(1+\xi^2)/(1+\xi^2+\xi^4)}$.
In static pure-AdS coordinates, $\Delta T$ is diagonal.  With
$t_M=\Delta T^M{}_M/\Ncal$ (no sum) and $t_\perp=t_y=t_w$, its components are
\begin{equation}
\begin{aligned}
 t_t&=1-q^{-1}-q\xi^6+\xi^5,
 &t_x&=1-q^{-1},\\
 t_\perp&=1-q,
 &t_z&=1+(\xi^4-1)q^{-1}-\xi^5.
\end{aligned}
\label{eq:zeroT-subtracted-source}
\end{equation}
This source is conserved in the background geometry and regular through
$\xi=1$; no additional gravitational boundary condition is imposed there.
Unlike the isotropic finite-temperature ansatz, its stationary response retains distinct
longitudinal and transverse components.  With $\kappa=8\pi G_5\Ncal$, we shall
use the following pure-AdS radial-gauge metric:
\begin{equation}
 ds^2=\frac{1}{z^2}\left[
 -(1+\kappa H_t)dt^2+(1+\kappa H_x)dx^2
 +(1+\kappa H_\perp)(dy^2+dw^2)+dz^2\right].
\label{eq:zeroT-metric}
\end{equation}
The metric functions are $H_i=H_i(\xi)$.
Define $\tau=H_t+H_x+2H_\perp$ and
$\vartheta=t_t+t_x+2t_\perp+t_z$.
The external Einstein equations and the radial constraint are
\begin{equation}
 \xi^2 H_i''-3\xi H_i'-\xi\tau'
 =-2\left(t_i-\frac{\vartheta}{3}\right),
 \qquad
 \xi\tau'=-\frac23t_z,
 \qquad i\in\{t,x,\perp\}.
\label{eq:zeroT-einstein}
\end{equation}
Primes in this subsection denote $\xi$ derivatives.  We fix the boundary
metric by $H_i(0)=0$ and select the solution with no added infrared-growing
homogeneous modes, $H_i=o(\xi^4)$ as $\xi\to\infty$, while imposing the
radial constraint.  The resulting solution has
$H_i=\cO(\xi^4\log\xi)$ near the boundary and
$H_t=\cO(\log\xi)$,
$H_x=-\xi/3+\cO(\log\xi)$,
$H_\perp=\xi/3+\cO(\log\xi)$ in the infrared.

For a pure-AdS strip with turning point $z_t$, let $\xi_t=\sqrt E\,z_t$ and
$\rho=(\xi/\xi_t)^6$.  Its dimensionless width is
$\Lambda=\sqrt E\,\ell
=2\sqrt\pi\,\Gamma(2/3)\xi_t/\Gamma(1/6)$.
For $\alpha\in\{\parallel,\perp\}$, including both radial halves,
the first-order area variation gives
\begin{equation}
\begin{aligned}
 \Delta S_\alpha^{(1)}
 &=\Ncal E V_2\,\mathcal F_\alpha(\Lambda),
 &\mathcal F_\alpha
 &=2\pi\int_0^{\xi_t}
 \frac{\mathcal W_\alpha(\xi,\rho)}{\xi^3\sqrt{1-\rho}}\,d\xi,\\
 \mathcal W_\parallel&=2H_\perp+\rho H_x,
 &\mathcal W_\perp&=H_x+(1+\rho)H_\perp .
\end{aligned}
\label{eq:zeroT-strip-area}
\end{equation}
The boundary behaviour makes the integral ultraviolet finite.  The
$\cO(\xi)$ infrared growth gives an integrable $\cO(\xi^{-2})$ tail, and the
near-turning-point region contributes at most $\cO(\xi_t^{-1})$.
Consequently, both $\mathcal F_\alpha$ approach finite constants as
$\Lambda\to\infty$: the selected stationary electric correction is area-like
and contains no term proportional to $\Ncal E^{3/2}V_2\ell$ at first order.

For small $\kappa$, this saturation is approached within the parametrically
wide regime $1\ll\sqrt E\,\ell\ll\kappa^{-1}$, where the metric perturbation
remains small.  The large-width limit above concerns the first-order
coefficient; it does not determine the non-linear infrared completion or
the asymptotic entropy scaling at fixed $\kappa$.

The result refers to the specified source and boundary conditions.  Further
isotropic source completions can change the common scalar response without
changing the orientation difference at this order, provided the tensor
boundary data are held fixed.  Such completions change normalisable state
data, not just the UV subtraction scheme.

\bibliographystyle{JHEP}
\bibliography{references}
\end{document}